  \documentclass[journal]{IEEEtran}
\usepackage{balance}
\usepackage{graphicx}
\usepackage [noadjust]{cite} 
\usepackage{bm}  
\usepackage{amsmath}
\usepackage{amssymb}
\usepackage{enumerate}
\usepackage{stfloats}
\usepackage{cases}
\usepackage{epstopdf}
\usepackage{soul,color} 
\usepackage{hyperref}
\usepackage[percent]{overpic}
\usepackage{tabularx}
\usepackage[table]{xcolor}
\usepackage{multirow}
\usepackage{booktabs}  
\usepackage{tabu} 

\usepackage{xr}    

\usepackage{textpos}

\usepackage{algorithm}
\usepackage{algpseudocode}

\newcommand{\calX}{{\mathcal{X}}}
\newcommand{\calXbold}{\boldsymbol{\mathcal{X}}}
\newcommand{\calH}{{\mathcal{H}}}
\newcommand{\calHtil}{\mathcal{\tilde{H}}}
\newcommand{\calHbold}{\boldsymbol{\mathcal{H}}}
\newcommand{\calF}{{\mathcal{F}}}

\newcommand{\calVbold}{\boldsymbol{\mathcal{V}}}
\newcommand{\calT}{{\mathcal{T}}}

\newcommand{\calS}{{\mathcal{S}}}
\newcommand{\calE}{{\mathcal{E}}}

\newcommand{\hbold}{\boldsymbol{h}}

\newcommand{\phibold}{\boldsymbol{\phi}}
\newcommand{\Phibold}{\boldsymbol{\Phi}}

\newcommand{\F}{\mathcal{F}}

\newcommand{\R}{\mathcal{R}}
\newcommand{\Rtil}{\mathcal{\tilde{R}}}

\newcommand{\HH}{\mathcal{H}}

\newcommand{\plexp}{\alpha}

\usepackage{amsthm}

\newtheorem{lemma}{Lemma}

\newtheorem{theorem}{Theorem}

\newtheorem {remark}{Remark}
\newtheorem {definition}{Definition}
\newtheorem {example}{Example}

\begin{document}
	\allowdisplaybreaks
	\title{
    Higher-Order Meta Distribution Reliability Analysis of Wireless  Networks}
	
  \author{Mehdi Monemi,~\IEEEmembership{Member,~IEEE,} Mehdi Rasti,~\IEEEmembership{Senior~Member,~IEEE,} S. Ali Mousavi,  Matti~Latva-aho,~\IEEEmembership{Fellow,~IEEE,} Martin Haenggi, \IEEEmembership{Fellow,~IEEE}
\thanks{\ 
Mehdi Monemi, Mehdi Rasti, and Matti Latva-aho are with the Centre
for Wireless Communications (CWC), University of Oulu, Oulu, Finland (emails: mehdi.monemi@oulu.fi, mehdi.rasti@oulu.fi, and matti.latva-aho@oulu.fi).

S. Ali Mousavi is with the Department of Electrical Engineering, Shiraz University of Technology, Shiraz, Iran (e-mail: al.mousavi@sutech.ac.ir).

Martin Haenggi is with the Department of Electrical Engineering, University of
Notre Dame, Notre Dame, IN 46556 USA (e-mail: mhaenggi@nd.edu).
}
\thanks{This work is supported by Business Finland via the 6GBridge - Local 6G project under Grant  8002/31/2022, Research Council of Finland via 6G Flagship under Grant 369116, and Research Council of Finland via the Profi6 Project under Grant 336449.}
}


\maketitle

\begin{abstract}
    Communication reliability, as defined by 3GPP, is the probability of achieving a desired quality of service (QoS). Traditionally, this metric is evaluated by averaging the QoS success indicator over spatiotemporal random variables. Recently, the meta distribution (MD) has emerged as a two-level analysis tool that characterizes system-level reliability as a function of link-level reliability thresholds. However, existing MD studies have two limitations. First, they focus exclusively on spatial and temporal randomness corresponding to node distribution and fading channels, respectively, leaving stochastic behaviors in other domains largely unexplored. Second, they are restricted to first-order MDs with two randomness levels, restricting applicability to scenarios requiring higher-order MD characterization. 
    To address these gaps, we propose a hierarchical framework for higher-order MD reliability in wireless networks, where each layer's success probability is formulated and fed into the next layer, yielding overall MD reliability at the highest level. We apply this framework to wireless networks by capturing three levels of temporal dynamics representing fast, slow, and static random elements, and provide a comprehensive second-order MD reliability analysis for two application scenarios. The effectiveness of the proposed approach is demonstrated via these representative scenarios, supported by detailed analytical and numerical evaluations. Our results highlight the value of hierarchical MD representations across multiple domains and reveal the significant influence of inner-layer target reliabilities on overall  performance.
\end{abstract}


\section{Introduction}

\begin{textblock*}{12cm}(6cm,-16.9cm)  
   Accepted to be published in IEEE Transactions on Wireless Communications \end{textblock*}

Reliable communication in wireless networks is essential for supporting a wide range of applications, from enhanced mobile broadband to ultra-reliable low-latency communication (URLLC). Traditional reliability analyses typically compute the success probability by averaging the quality-of-service (QoS) success indicator over all sources of randomness in the system. Although straightforward, this approach lacks the ability to capture the hierarchical dependencies of various random elements in the overall performance reliability. In particular, it does not quantify how link-level reliability influences overall system-level reliability. For example, a network may have an average link success probability of 95\%, but this single value can hide large variations across network deployments: for some deployments the link success probability might exceed 99\%, while for others it could be closer to 91\%. Specifically, it cannot quantify the minimum desired link reliability across all possible network deployments. Such a valuable insight is obscured when taking the expectation jointly over all random elements in a single step, without distinguishing between their roles and temporal dynamics hierarchy in the system.  
To address this limitation, the meta distribution (MD) framework adopts a hierarchical approach to calculating success probabilities \cite{9389803,9389789}. By partitioning the collection of all random elements into ordered classes $\calX_0$ and $\calX_1$, the first-order MD reliability is defined as the probability that the conditional QoS success probability exceeds a target threshold, i.e., 
$
\mathbb{P}_{\calX_1}\big(\mathbb{P}_{\calX_0}(Q > q \mid \calX_1) > p_1\big),
$ where $Q$ quantifies the QoS, and \(p_1\in[0,1]\) and $q>0$ are the target thresholds. This differs from the conventional reliability metric $\mathbb{P}(Q>q)$ which averages the QoS success indicator over all random elements simultaneously. Consequently, the MD reliability provides a more nuanced understanding of how success probability at inner layer affects the overall network performance.

MD-based analyses have been leveraged for wireless networks in many of the existing works. 
In the context of performance evaluation and reliability analysis, several works have investigated the calculation of the signal-to-interference (SIR) or signal-to-interference-plus-noise (SINR) meta distribution. 
The SIR MD for Poisson network models was initially introduced and evaluated in \cite{7345601}. Subsequent research extended the results to various device-to-device (D2D) and cellular networks \cite{7893755,8115171,8587178}. In addition to SIR, several studies have investigated the SINR MD for Poisson network models. 
For example, in \cite{8290712}, the MD of the secrecy rate of a single node in the presence of randomly located eavesdroppers was investigated.
In \cite{9982594} the MD of the downlink rate of the typical UAV under base station (BS) cooperation
in a cellular-connected UAV network was studied using a standard beta distribution approximation. 
The authors of \cite{10659172} have investigated the rate MD in ultra-reliable low-latency communication (URLLC) D2D networks considering the errors due to the misalignment of radiated beams.
In \cite{8474350}, the energy and rate MD have been leveraged
 to quantify a performance metric termed {\it wirelessly powered spatial transmission efficiency} for D2D networks. By formally characterizing the link and spatial reliability concepts and utilizing MD reliability analysis, the authors of \cite{10142008} have derived closed-form expressions for bandwidth requirements needed for guaranteeing target values of link and spatial reliability in URLLC networks.  Following similar strategies for providing link and spatial reliabilities, \cite{10078093} underscored the substantial bandwidth demands, reaching on the order of gigahertz, to effectively support URLLC in future wireless networks. To address these considerable requirements, the authors suggested network densification and multi-connectivity as key mitigation strategies. 


While all the aforementioned works on MD analysis are limited to Poisson network models, the study of MD in wireless networks is not restricted to such models. 
Given the difficulty in analyzing non-Poisson network models, the authors of \cite{8648502} proposed a simplified scheme called AMAPPP (``Approximate meta distribution analysis using the PPP") to approximate the SIR MD for non-Poisson networks. Considering a clustering strategy for wireless devices around the access points, the authors of \cite{9913703} formulated the outage probabilities of transferred energy and transmitted rate conditioned on the locations of network devices, and then they calculated the MDs to investigate the average proportion of the wireless devices in one cluster that achieves successful performance in terms of energy transfer and transmission rate while satisfying the reliability constraint. 


To the best of our knowledge, all existing works in the literature investigating the MDs in wireless networks have focused on {\it first-order} spatiotemporal MD analysis, considering the random {\it spatial} distribution of the wireless nodes and the {\it temporal} characteristics of the small-scale fading channels.  
Among existing studies, the work in \cite{9389789} is the only one that presents a compact representation of higher-order MDs, although its primary focus remains on first-order MD. However, the proposed higher-order representation is purely mathematical, offering no insight or  examples regarding its application to wireless networks. Furthermore, it does not explore the relationship between MDs of different orders. To address this gap, in this work we develop a  hierarchical representation of higher-order MDs and provide a framework for applying it to the analysis of MD-reliability of wireless networks.    
The main contributions of this work are listed as follows:

\begin{itemize}
    \item 
    We formally express the zeroth-order (non-MD) and first-order MD reliability representation and provide examples of wireless applications where MD reliability characterization extends beyond the conventional spatiotemporal domain that has been widely explored and discussed in the literature.
   
    \item Building on the strengths of first-order MD, 
    we introduce the higher-order MD reliability representation where the random variables are partitioned into multiple ordered classes and the reliability analysis is 
    conducted hierarchically across several domains. 
    Specifically, we propose a framework for analyzing higher-order MD reliability in wireless networks by considering three levels of temporal dynamics,  
    including fast time-varying, slowly time-varying, and static random elements. 
    The MD at each layer is explicitly formulated and characterized to be leveraged at the higher layer, where the ultimate MD reliability measure is  obtained at the highest layer.
    
    \item As a first application of the higher-order MD reliability analysis, we focus on a canonical three-layer stochastic geometry model, where the bottom and top layers correspond to time-varying fading channels and the static point process of BSs, respectively. The middle layer captures slow time-varying randomness, where 
    the set of interfering BSs is modeled as a randomly thinned point process selected as a subset of all BSs. 
   We derive closed-form analytical expressions for the MD reliability and illustrate, through numerical results, the insights revealed by second-order MD compared to first-order MD, and non-MD analyses.

    \item 
     In the second application, we study the {\it temporal-spectral-spatial} MD reliability in wideband frequency-hopping spread spectrum (FHSS) THz networks. 
    Our study gives an important understanding about the interplay between target threshold values on the MD reliability and     provides insight into balancing spectrum allocation to achieve optimal spatial MD reliability while meeting temporal and spectral reliability targets.
\end{itemize}


The remainder of the paper is structured as follows. Section II  investigates the conventional (non-MD) and first-order MD reliability analysis and provides examples in wireless applications where first-order spatiotemporal and non-spatiotemporal MD reliability analysis can be leveraged. Section III extends the MD reliability characterization for higher-order MDs and provides a framework for analyzing the higher-order MD reliability analysis of wireless networks. Section IV investigates the analysis of the second-order MD reliability for two different applications and provides supporting discussions for each example. Finally, the paper is concluded in Section V.

\section{Non-MD and First-Order MD-based  Reliability Analysis}
\label{sec:II}
In this section, we study the conventional non-MD reliability as well as first-order MD reliability in wireless networks. 
\subsection{Conventional (non-MD) Reliability }

The communication reliability, as defined by 3GPP \cite{3gpp_tr_38913_18}, refers to the success probability of delivering $l$ bits with a time delay lower than a user-plane deadline threshold $t_{\mathrm{th}}$. Although primarily introduced for low-latency services (such as URLLC), it applies to different network services including URLLC, enhanced mobile broadband (eMBB), and massive machine-type communication (mMTC). This definition can further be generalized as follows to encompass a broader range of applications:
\begin{definition}
    The reliability measure $R$ is the probability that the QoS measure function $Q$ be higher than a minimum required threshold $q$, i.e.,
    \begin{align}
        \label{eq:R}
        R(q)=\mathbb{P}_{\boldsymbol{\calX}}(Q>q),
    \end{align}
    where $\boldsymbol{\calX}$ is the collection of random elements including {\it temporal} random variables (e.g., small-scale fading), {\it spatial} random variables (if any, such as the stochastic point process corresponding to the positions of users/BSs), or any additional random variables. 
    The QoS function $Q$ may take different forms depending on the service type and system model.
\end{definition}
\begin{example}
    \label{ex:1D}
    {\it Conventional stochastic geometry based reliability analysis for delay-tolerant  services \cite{7893755,8795532,7733098}:
    }
    {\rm
    Consider a downlink communication scenario where BSs are randomly scattered in the network region according to a stationary Poisson point process (PPP). Each user is provided with some delay-constrained service through the nearest BS with packets of $l$ bits at time duration $t_l$ obtained from the Shannon-Hartley capacity. We have $t_l(\mathrm{SINR}(\calHbold,\Phibold))=l/(W\log(1+\mathrm{SINR}(\calHbold,\Phibold))$, where $\calHbold$ and $\Phibold$ are the random variables corresponding to small-scale fading channels and the point process relating to the BSs locations respectively, and $W$ is the bandwidth. The reliability is obtained as $R=\mathbb{P}(1/t_l(\mathrm{SINR}(\calHbold,\Phibold))>1/t_{\mathrm{th}})$.
    For the simple case of orthogonal frequency carriers where the interference is negligible relative to noise through coordinating the frequency resources in nearby cells, and considering that all links follow same channel fading statistics, the function $\mathrm{SINR}(\calHbold,\Phibold)$ can be replaced by the simpler signal-to-noise-ratio (SNR) function $\mathrm{SNR}(\HH,\R)$ where $\calH$ is the scalar small-scale fading of the typical link and $\R$ is the length of the typical link. Considering the independence of the spatial and temporal distributions, the reliability is then obtained as $\iint_{(h,r) \in \calS} f_\HH(h) f_\R(r) \, dh \, dr$, where $f_{\HH}(h)$    is the probability density function (pdf) of the fading channel for each of the users,  $f_\R(r)=2\pi\lambda re^{-\lambda \pi r^2}$ is the pdf of the distance $\R$, $\lambda$ is the intensity of the PPP, and finally $\calS$    is the region of interest characterized as $\calS=\{(h,r)\in \mathbb{R}_{+}^{2}\mid t_l(\mathrm{SNR}(h,r))\leq t_{\mathrm{th}}\}$. Here, the SNR function can be modeled as $\mathrm{SNR}(h,r)=\frac{P_{\mathrm{T}}  G_{\mathrm{T}} G_{\mathrm{R}}c^2}{(4\pi f)^2} \left( \frac{ h r^{-\plexp}}{N_0 W}\right)$, where $c$ is the speed of light, $f$ is the frequency, $\plexp$ is the path loss exponent, $W$ is the bandwidth, $N_0$ is the spectral density of the noise, $P_{\mathrm{T}}$ is the transmit power, and $G_{\mathrm{T}}$ and $G_{\mathrm{R}}$ are the transmit and receive antenna gains, respectively.
    }
\end{example}

\subsection{First-Order Spatiotemporal MD Reliability}
To provide a hierarchical reliability analysis, 
the first-order MD reliability is defined as follows:
\begin{definition}
Assume that the collection of random variables $\calXbold$ is partitioned into the ordered classes $\calX_0$ and $\calX_1$. Given the two parameters $q$ and $p_1\in[0,1]$, the (first-order) MD reliability measure is defined as\footnote{-While the subscripts of $\mathbb{P}$ in \eqref{eq:2O_MD} are technically redundant, we retain them for enhanced clarity. This holds for the subscript of $\mathbb{P}$ in \eqref{eq:R} as well.}
\begin{align}
\label{eq:2O_MD}
    R_{[1]}(p_1;q)=\mathbb{P}_{\calX_1}
(\mathbb{P}_{\calX_0}(Q>q\mid\calX_1) > p_1),    
\end{align}
where  $p_1$ is the first-level target reliability value. 
\end{definition}

 From \eqref{eq:2O_MD} it is seen that $R_{[1]}(p_1;q)$ measures the probability of achieving the desired QoS conditioned on $\calX_1$ be higher than a threshold value $p_1$. 
  For now, consider that $\calX_0$ and $\calX_1$ correspond to temporal and spatial random variables, respectively. Assuming $\calX_1$ to be an ergodic process, the MD reliability $R_{[1]}(p_1;q)$ captures the overall {\it spatial reliability} over the service region by guaranteeing the {\it link reliability} threshold of $p_1$ over all realizations of spatial variables (e.g., locations of the users or BSs).
 The following example presents a foundational system model that serves as the basis for reliability analysis conducted in many studies investigating the reliability of wireless communications using the MD approach.

\begin{example} {\it First-order MD reliability for delay-tolerant  services \cite{10142008,10659172}:}
    \label{ex:2D_spatial_temporal}
    {\rm
    Consider the network service expressed in Example \ref{ex:1D}. Letting $\calX_0=\calHbold$ and $\calX_1=\Phibold$, the MD reliability $R_{[1]}(p_1;q)$  yields the fraction
of links in all realizations of the point process that achieve $\mathrm{SINR}>q$ with probability $p_1$.
}

\end{example}

The study of MD reliability in the spatiotemporal domains is not limited to delay-tolerant (e.g., URLLC) and rate-tolerant (e.g., eMBB) services, as exemplified in the following.
\begin{example} {\it First-order MD reliability for the harvested energy analysis \cite{8474350}:}
    \label{ex:2D_spatial_temporal2}
    {\rm
    Consider a collection of D2D devices scattered in the network with a spatial distribution described by some point process. The QoS function can be considered as the amount of harvested energy during each time slot, denoted by $\calE$, which can be formulated as a function of fading channels $\calHbold$ and users' positions corresponding to $\Phibold$ \cite{8474350}. The reliability measure $R_{[1]}$ represents the MD of the harvested energy $\calE(\hbold,\phibold)$ guaranteeing the link energy success probability higher than the threshold $p_1$ conditioned on spatial positions of users and RF transmitters. This problem follows a spatiotemporal MD analysis similar to Example \ref{ex:2D_spatial_temporal}.
    }
\end{example}
\subsection{First-Order Non-Spatiotemporal MD Reliability}
\label{sec:first_order_non_spatiotemporal}
As previously mentioned, existing studies on first-order MD reliability primarily focus on spatiotemporal structures, where the inner and outer layers correspond to time and space domains, respectively. However, many practical scenarios do not naturally conform to this framework, yet can still benefit from MD-based reliability analysis using the formulation in \eqref{eq:2O_MD}. The application of MD reliability to such non-spatiotemporal configurations remains unexplored in the literature. In what follows, we present a simple illustrative example.

  \begin{example}
  \label{ex:nonspatio1}
       {\it End-to-end link reliability leveraging the MD of radio-link and fronthaul/backhaul connections:}
       {\rm
       Consider a URLLC network service wherein an end-to-end connection is set between a fixed user and the associated access point. 
       The end-to-end delay can be modeled as $t=t_l(\mathrm{SINR}(\calHbold,\Phibold_0))+\calT$, where $t_l$ is the radio link delay corresponding to the transmission of the packet of $l$ bits from the user to the access point (e.g., gNodeB), $\Phi_0$ is the set of locations of the network nodes, which are assumed to be fixed, and $\calHbold$ denotes the small-scale fading channel of the links, and  $\calT$ is the additional delay due to queuing, routing, processing, etc., in the fronthaul/backhaul of the network, relating to the connection from the access point to the final destination (e.g., user plane function). By considering the radio link reliability of $p_1$, and assuming a statistical model for $\calT$, the overall MD reliability is calculated according to \eqref{eq:2O_MD} where $\calXbold=\{\calHbold,\calT\}$, in which $\calHbold\equiv \calX_0$ and $\calT\equiv \calX_1$. For the case of orthogonal multiple access where no interference is imposed from other links, similar to Example 1, the function $\mathrm{SINR}(\calHbold,\Phibold_0)$ reduces to $\mathrm{SNR}(\calH;R_0)$ where $\calH\in\calHbold$ is the scalar small-scale fading of the intended communication link, and $R_0$ is the distance of the link which is assumed to be a fixed here. Given $t_{\mathrm{th}}$ and $p_1$, the MD reliability is obtained as
       \begin{align}
            \label{eq:2O_MD372}
                R_{[1]}&=\mathbb{P}_{\calT}
            (\mathbb{P}_{\calH}(t<t_{\mathrm{th}}\mid\calT) > p_1)
            \notag
            \\
            &=\mathbb{P}_{\calT}
            (\mathbb{P}_{\calH}(t_l(\mathrm{SNR}(\calH;R_0))+\calT<t_{\mathrm{th}}\mid\calT) > p_1)
            \notag
            \\
            &=
            \bar{F}_{\bar{F}_{\calH}(\mathrm{SNR}^{-1}(t_l^{-1}(t_{\mathrm{th}}-\calT);R_0)\mid \calT)}(p_1), 
       \end{align}
       where $\mathrm{SNR}^{-1}(\gamma; R_0) = \{ h \mid \mathrm{SNR}(h; R_0) = \gamma \}$ and
       $\bar{F}_X$ denotes the complementary cumulative distribution function (ccdf) of $X$.
       Note that the small-scale fading random variable $\calH$ and the random delay process $\calT$ corresponding to the fronthaul/backhaul transmission are both temporal random variables, considered uncorrelated in most practical scenarios.
       }
    \label{ex:2D_non_spatial_temporal}
  \end{example}
\section{Beyond First-Order MD Reliability Analyses}
\label{sec:beyond_first_order}
 Most studies in the literature use a first-order MD reliability framework with spatiotemporal decomposition as exemplified in Examples \ref{ex:2D_spatial_temporal} and \ref{ex:2D_spatial_temporal2}. However, the MD’s applicability in wireless network reliability extends beyond this. 
    Building on the strengths of first-order MD analyses of the reliability over space and time domains, we extend this to a broader, higher-order MD reliability analysis over various domains. 
    This allows for a more nuanced understanding of reliability across different dimensions. For instance, higher-order MD analyses can capture complex interactions between factors like signal strength variations, delay jitter, fading, frequency statistics, and packet loss variations. By analyzing these dependencies in a hierarchical structure, we can gain valuable insights into resource allocation strategies and improve network performance prediction, leading to more robust and reliable wireless networks, in the sense that the impact of a change in the reliability measure at each dimension can be accurately monitored and explored in the overall reliability of the system. 

Formally, higher-order MDs are defined as follows:
\begin{definition}
\label{def:hierarchical_md}
    Let $Q$ be a function of random elements $\boldsymbol{\calX}$, which are partitioned into the ordered classes $\calX_0, ..., \calX_n$. Let the random variables $P_1^{(n)}, ..., P_n^{(n)}$ iteratively be
\begin{align}
     P_1^{(n)}(q) &\triangleq \mathbb{P}_{\calX_0}(Q > q) = \mathbb{P}(Q > q \mid \calX_1,  ..., \calX_n) 
    \notag \\
    P_2^{(n)}(p_1;q) &\triangleq \mathbb{P}_{\calX_1}(P_1^{(n)} > p_1) = \mathbb{P}(P_1^{(n)} > p_1 \mid \calX_2,  ..., \calX_n)
    \notag \\
    &\vdots 
    \notag \\
\label{eq:MDitterative}
    P_n^{(n)}(\boldsymbol{p}_{n-1};q) &\triangleq \mathbb{P}_{\calX_{n-1}}(P_{n-1}^{(n)} > p_{n-1}) = \mathbb{P}(P_{n-1}^{(n)} > p_{n-1} | \calX_n),
\end{align}
where $\boldsymbol{p}_k \triangleq (p_1, ..., p_k) \in [0, 1]^k$. The $k$-th order MD (also referred to as $k$-th order MD reliability) denoted by $R_{[k]}^{(n)}$ is defined as
\begin{align}
     R_{[0]}^{(n)}(q)&\triangleq \mathbb{P} (Q>q),
     \notag \\
\label{eq:MDitterative_rel}
    R_{[k]}^{(n)}(\boldsymbol{p}_k; q) & \triangleq \mathbb{P}(P_k^{(n)} > p_k), \quad k \in [n],
\end{align}
If $k=0$, this is the standard ccdf of $Q$ corresponding to the conventional reliability $R(q)$ in \eqref{eq:R}. If $k=n$, the MD is {\it maximally discriminative},  since the decomposition of $\boldsymbol{\calX}$ into $n+1$ classes is fully exploited. In contrast, for $k < n$, only $k+1$ classes are taken into account since $\calX_k,...,\calX_n$ are lumped together and expected over in the last step of calculating $\mathbb{P}(P_k^{(n)}>p_k)$. We write $R_{[n]}$ for simplicity for the maximally discriminative MD.
It can be expressed compactly as
\begin{multline}
\label{eq:Rncompact}
    R_{[n]}(\boldsymbol{p}_n,q) \triangleq \mathbb{P}(P_n^{(n)} > p_n)=\\
     \mathbb{P}_{\calX_n} \left( \mathbb{P}_{\calX_{n-1}} \left( ... \left( \mathbb{P}_{\calX_1} (Q > q) > p_1 \right) > ... > p_{n-1} \right) > p_n \right).
\end{multline}
\end{definition}
%
\begin{remark}
\label{rmk:MD_order_relation}
    Since $\mathbb{E}(P_k^{(n)}) = \mathbb{P}(P_{k-1}^{(n)} > p_{k-1})$ the MDs for $k=2,...,n$ are related as
    \begin{equation}
        \label{eq:MD_iteratively_related}
        R_{[k-1]}^{(n)}(\boldsymbol{p}_{k-1}, q) = \int_0^1 R_{[k]}^{(n)}(\boldsymbol{p}_k, q) dp_k.
    \end{equation}
\end{remark}
\begin{remark}
Note that removing the outermost layer of $R_{[n]}$ in \eqref{eq:Rncompact} which results in
 $\mathbb{P}_{\calX_{n-1}} \left( ... \left( \mathbb{P}_{\calX_1} (Q > q) > p_1 \right) > ... > p_{n-1} \right)$ does not yield the $(n-1)$-th order MD since it is a function of $\calX_n$.
\end{remark}
\begin{remark}
\label{rmk:compact_vs_ours}
   A compact form of higher-order MD representation was introduced in \cite{9389789} where in the context of MD reliability, that definition is translated into the representation of $R_{[n]}(\boldsymbol{p}_{n}, q)$ expressed in \eqref{eq:Rncompact}. 
   In Definition \ref{def:hierarchical_md}, we have presented a hierarchical form of calculating $n$-th order MD in \eqref{eq:MDitterative} and \eqref{eq:MDitterative_rel}, where MDs of different orders are iteratively related according to \eqref{eq:MD_iteratively_related}.  This approach not only enables the presentation of the relation between MDs of different orders, but also provides a more structured way to follow corresponding mathematical calculations for obtaining the overall MD reliability leveraging the successive calculation of random variables $P_k^{(n)},k\in\{1,\cdots, n\}$, followed by computing $R_{[n]}$ in the uppermost layer as will be shown in the analysis of two applications presented in the next section.
\end{remark}
\begin{figure}
    \centering
    \includegraphics[width=1.1\linewidth]{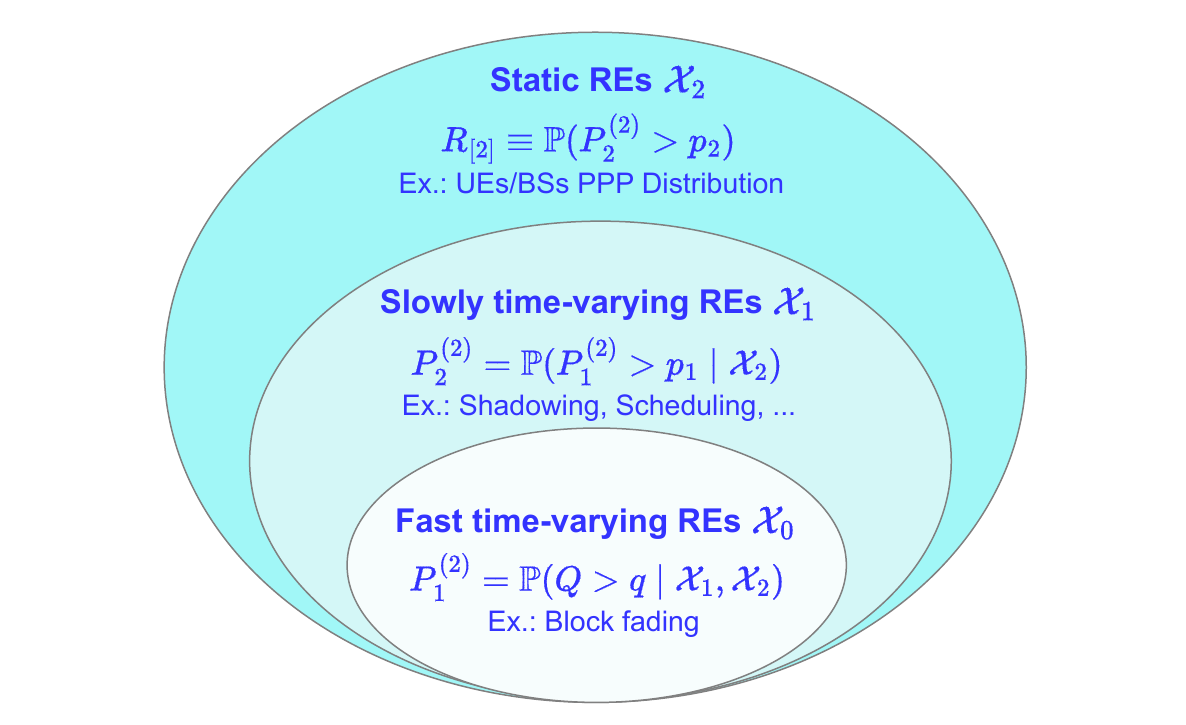}
    \caption{A  framework for higher-order ($n=2$) MD reliability analysis in wireless networks, where the random elements (REs) are partitioned into three ordered classes with different levels of temporal dynamism. }
    \label{fig:higher-struct}
\end{figure}

The higher-order MD reliability analysis can be applied to wireless networks involving random elements with varying levels of temporal dynamics. Fig.~\ref{fig:higher-struct} illustrates such a system, structured into three levels of dynamicity.  At the lowest level, corresponding to the highest dynamicity, the random variable $P_1^{(2)}$ is calculated as the QoS success probability conditioned on all random elements except the fast time-varying ones, mainly attributed to the channels (block) fading. The middle layer accounts for calculating $P_2^{(2)}$, which captures the effects of slowly time-varying random elements. Depending on the system model and the network under study, these may involve large-scale shadowing, random user mobility, scheduling and stochastic channel access mechanisms,  service time variations in network core and edge devices, and other network dynamics. 
Finally, the MD reliability  $R_{[2]}$ corresponding to the second-order {\it MD coverage probability}  is derived in the top layer, where the expectation is taken with respect to the random elements governing the locations of users and BSs.

\begin{remark}
Although Fig.~\ref{fig:higher-struct} illustrates a three-layer second-order MD reliability framework for wireless networks, this approach can be generalized to higher-order MDs. To do so, the random elements in the middle layer can be further partitioned into multiple ordered classes based on their temporal dynamics.  For example, elements with slower temporal variations, such as large-scale shadowing, can be associated with a lower-order class, while elements with faster dynamics, such as those relating to random/pseudorandom channel access, may be assigned to a higher-order one.
\end{remark}

\begin{example}
    \label{ex:mobility_aware}
    {\it Application of MD to assess mobility-aware system reliability: }
{\rm
     Mobility can significantly impact system reliability by introducing challenges like Doppler shifts and handoff delays. In practice, different users might have different velocities at different snapshots of time and spatial realizations of the point process, and thus the mobility can be modeled as a random variable whose distribution may be modeled as uniform or Gaussian \cite{8673556}.
    Consider a network similar to Example \ref{ex:1D}, where we model the velocity of users as random variables denoted by $\calVbold$. The QoS function can be expressed as $Q(\calVbold,\calHbold,\Phibold) = \delta(\calVbold) / t_l(\text{SINR}(\calHbold,\Phibold))$, where $\delta(\calVbold) \leq 1$ captures the capacity degradation due to mobility-related issues, such as Doppler spread and the reduction of the channel coherence time. 
   Here, $\mathcal{X}_0 \equiv \calHbold$, $\mathcal{X}_1 \equiv \calVbold$, and $\mathcal{X}_2 \equiv \mathbf{\Phi}$ denote random variables represented in the temporal, spatiotemporal, and spatial domains, corresponding to fast time-varying, slowly time-varying, and static variations, respectively. From \eqref{eq:MDitterative} and \eqref{eq:MDitterative_rel}, the second-order MD reliability is obtained by calculating  $P_1^{(2)}=\mathbb{P}
     (
        Q(\calHbold,\calVbold,\Phibold) >q \mid \Phibold,\calVbold
    )
    $, and then  $ P_2^{(2)} =\mathbb{P}(P_1^{(2)}> p_1\mid \Phibold)
    $, 
    and finally $
     R_{[2]} =\mathbb{P}(P_2^{(2)}> p_2)$.
    }
\end{example}

The success probability thresholds values considered across inner layers characterize the MD reliability calculated at the outermost layer. For instance in the aforementioned example, the middle layer, conditioned on network deployment, measures the probability $P_2^{(2)}$ (expected over user velocities) that the link-level QoS requirement $q$ is satisfied, compared against the mobility success threshold rate $p_1$ (e.g., with $p_1 = 0.9$ and uniform velocity distribution, corresponding to a requirement that the links in at least $90\%$ of velocities meet $q$). The outer layer then quantifies the fraction of deployments in which this mobility requirement is satisfied, relative to a system-level threshold $p_2$ (e.g.,  with $p_2 = 0.85$,  requiring that it holds in at least $85\%$ of network deployments). The resulting second-order MD reliability $R_{[2]}$ therefore captures the probability (over deployments) that the stated link-level and mobility-level requirements are satisfied. This provides network operators with the ability to monitor per-layer reliability guarantees, which conventional single-layer metrics cannot capture.

\begin{remark}
\label{rmk:practical}
    For practical or theoretical application scenarios where tractable closed-form expressions are unavailable, the overall MD reliability can be evaluated using numerical methods. Three approaches are possible: (i) pure numerical conditional integration, (ii) MD-based Monte Carlo simulation, as exemplified in the next section, and (iii) a hybrid method that combines the two, employing numerical integration in the lower layers and MD-based Monte Carlo in the outer layers.
\end{remark}

\section{Applications of Higher-Order MD Reliability in Wireless networks}
Building upon the three-layer structure proposed in the previous section, we now study two applications leveraging higher-order MD reliability in wireless networks.
\subsection{Second-Order MD-Reliability for Slowly Time-Varying Random Interfering BSs}
\label{sec:case_study_1}
In this part, we describe a canonical system setup in which the interfering and non-interfering BSs vary as slowly time-varying random variables in the middle layer of the proposed framework in Fig. \ref{fig:higher-struct}. We present the second-order MD reliability analysis and provide discussions on the numerical results. \subsubsection{System Model}
\begin{figure}
    \centering
\includegraphics[width=1\linewidth]{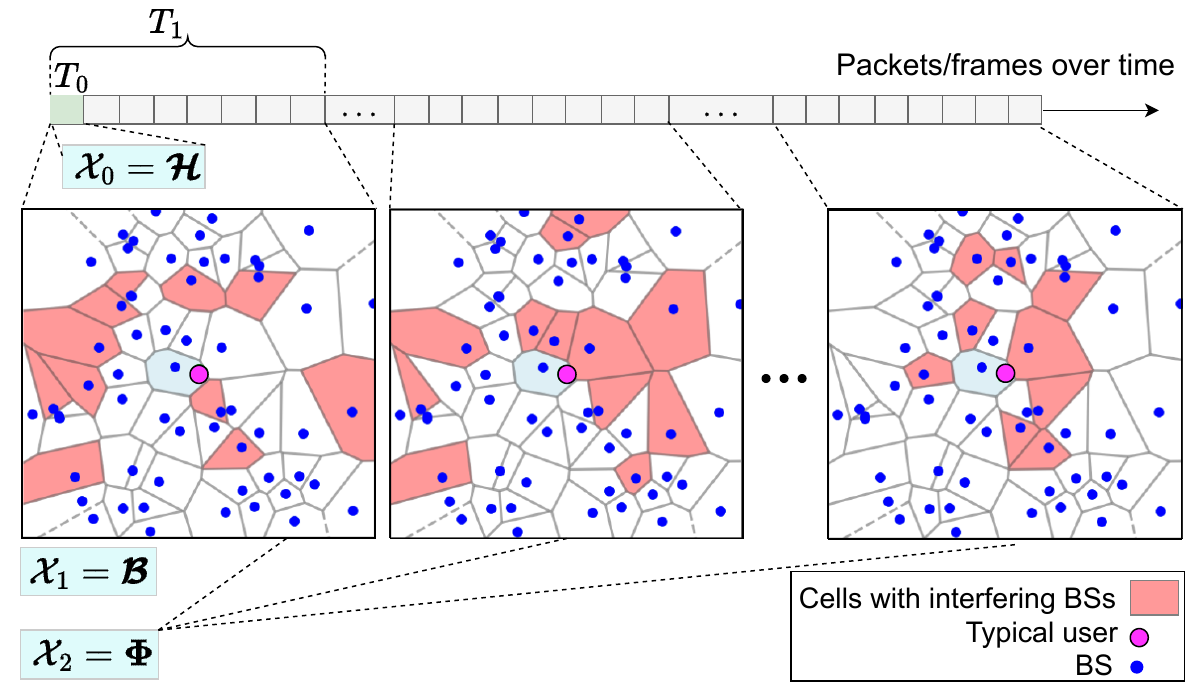}
    \caption{Three-layer wireless network with slowly time-varying interfering BSs used for second-order MD reliability analysis. Pink-filled cells denote interfering BSs. \(T_0\) and \(T_1\) indicate intervals over which \(\calHbold\) and \(\boldsymbol{\mathcal{B}}\) remain unchanged, respectively.}
    \label{fig:aaaaa}
\end{figure}

Consider a downlink cellular network wherein the spatial deployment of the BSs is modeled according to a homogeneous PPP $\Phibold \subset \mathbb{R}^2$ of density $\lambda$. BSs and users are equipped with omni-directional antennas. Each user associates with the nearest BS and all BSs are assumed to transmit at the same power level.  In our model, all channels experience power-law path loss with exponent $\plexp>2$ and are subject to i.i.d. Rayleigh fading $\calH_i\sim \mathrm{Exp}(1),\forall i$. 
We consider an interference-limited network where all BSs are continuously transmitting. Assume that each non-serving BS is an interferer with probability $\zeta\in (0,1]$. Considering that the typical user is served by the first (closest) BS, the interfering BSs are determined by $\boldsymbol{\mathcal{B}}=(b_i)$, where $(b_i)_{i>1}$ are i.i.d. Bernoulli with mean $\zeta$
. The set of interfering BSs $\boldsymbol{\mathcal{B}}$ is assumed to be unchanged for an interval $T_1 \gg T_0$, where $T_0$ is the coherence time of the block fading, and then replaced by an independent set. Similar to the block fading model, this interference structure can be viewed as a {\it block ALOHA} mechanism operating at the timescale of $T_1$. We adopt the set of fast time-varying, slowly time-varying and static random elements relating to the higher-order MD structure proposed in Fig. \ref{fig:higher-struct} as $\calX_0=\calHbold$, $\calX_1=\boldsymbol{\mathcal{B}}$ and $\calX_2=\Phibold$ respectively. Fig. \ref{fig:aaaaa} illustrates the system setup, depicting a realization of $\Phibold$ along with multiple corresponding realizations of $\boldsymbol{\mathcal{B}}$. The interfering BSs are the ones corresponding to pink-filled cells.
\subsubsection{Calculation of the second-order spatial MD reliability}
We are interested in calculating the second-order spatial MD reliability $R_{[2]}$ for guaranteeing SIR QoS threshold $q$ and target reliabilities $p_1$ and $p_2$ in the middle and top layers respectively. 
Let $\R_i$ and $\Rtil_i$ be the distance of the typical user to the $i$-th nearest BS and $i$-th nearest interfering BS. It is evident that \( \tilde{\mathcal{R}}_1 > \mathcal{R}_1 \). Besides, noting that conditioned on $\Phibold$, the interfering BSs are assigned independently at random with probability \( \zeta \in (0,1] \), it follows that \( \tilde{\mathcal{R}}_i > \mathcal{R}_i \) for all \( i \geq 1 \). In the special case where \( \zeta = 1 \), all non-serving BSs are interfering, and we have \( \tilde{\mathcal{R}}_i = \mathcal{R}_{i+1} \) for all \( i \geq 1 \).


The SIR of the typical user is considered as the QoS function $Q$, given by
\begin{align}
\label{eq:SIR44}
    Q\equiv\mathrm{SIR}
    =
    \frac{\calH_1 \R_1^{-\plexp}}{\sum_{i=1}^{\infty} \calHtil_i \Rtil_i^{-\plexp}}
    =
    \frac{\calH_1 \R_1^{-\plexp}}{\sum_{i=2}^{\infty} b_i\calH_i \R_i^{-\plexp}}
\end{align}
The desired QoS threshold is so $q=\left( 2^{\frac{l}{W  t_{\mathrm{th}} }} -1\right)$. 
Two scenarios will be investigated, {\it single-interferer} and {\it multi-interferer}. The latter accounts for considering all interfering BSs according to the stated probabilistic pattern, while the former considers only the first (i.e., strongest) interfering BS, corresponding to 
$\tilde{\R}_1$. While the multi-interferer analysis offers a more precise characterization of the MD reliability, the single-interferer case provides a highly tractable lower bound for the solution.

\begin{theorem}
\label{th:1}
    The second-order MD reliability for the single-interferer scenario is given by
    \begin{align}
     \label{eq:thdskf3}
     R_{[2]}(p_1,p_2;q)=
      \begin{cases}
     1-\left( 1-
        \frac{1}{\hat{p}_1^2}
        \right)^{\left\lfloor \frac{\ln (p_2)}{\ln(1-\zeta)}\right\rfloor+1 }, & \mathrm{if}\ \hat{p}_1 > 1
     \\
     1, & \mathrm{if }\ \hat{p}_1 \leq 1,
 \end{cases}
    \end{align}
    where $\lfloor x\rfloor$ denotes the floor function representing the greatest integer less than or equal to $x$, and
\begin{align}
\label{eq:p1hat}
    \hat{p}_1
    =
    \left[(p_1{q})\big/{\left(1-p_1\right)}\right]^{\frac{1}{\plexp}}
\end{align}
\end{theorem}
\begin{IEEEproof}
 See \ref{sec:proof_th1}.
\end{IEEEproof}

The following lemma supports the next theorem.
\begin{lemma}
    \label{lm:fgsk46}
    The expression $\mathbb{E}\left[\sum_{i=1}^{\infty}({ \Rtil_1}/{\Rtil_i})^{\plexp}\right]$ can be tightly approximated by $\frac{1+\delta\zeta}{1-\delta}$, where $\delta=2/\alpha$. 
\end{lemma}
\begin{IEEEproof}
    See \ref{sec:proof_lm1}.
\end{IEEEproof}
%
\begin{theorem}
\label{th:2}
    The second-order MD reliability for the multi-interferer scenario is tightly approximated by
     \begin{multline}
        \label{eq:hfdsnxmmzevd}
         {R_{[2]}}(p_1,p_2;q)
         \approx 
         \\
        \begin{cases}
        1-\left( 1-\frac{1}{\hat{p}_1^2}
        \left(\frac{1+\delta\zeta}{1-\delta}\right)^{\delta} \right)^{\left\lfloor \frac{\ln (p_2)}{\ln(1-\zeta)}\right\rfloor+1 } 
        & 
        \mathrm{if}\ \hat{p}_1 > \left(\frac{1-\delta}{1+\delta\zeta}\right)^{\delta/2}
        \\
        1, & \mathrm{else }
    \end{cases}
    \end{multline}
\end{theorem}
\begin{IEEEproof}
    See \ref{sec:proof_th2}.
\end{IEEEproof}
\subsubsection{Numerical Results and Discussion}
\label{sec:case_study_1_numerical}
   
\begin{algorithm} 
    \caption{Second-Order MD Reliability Monte Carlo Simulation for  \eqref{eq:thdskf3} and \eqref{eq:hfdsnxmmzevd}}
    \begin{algorithmic}[1]
    \Statex {\bf \hspace{-15pt} Initialization:}

\State Set the number of trials: \( N_0 \) (inner loop), \( N_1 \) (middle loop), \( N_2 \) (outer loop);

\State Define \( S_0 \), \( S_1 \), and \( S_2 \) as conditional success counters for the inner, middle, and outer loops, respectively;

\State Set $S_2\leftarrow 0$; 

\Statex {\bf \hspace{-15pt} Main Procedure:}

\For{\( i_2 = 1 \) to \( N_2 \)}

    \State Sample \( \calX_{2,i_2} \) from \( \calX_2\equiv\boldsymbol{\Phi} \), where $ \boldsymbol{\Phi} $ is a PPP with density $\lambda$ representing BSs' locations;
    
    \State Sort BSs by distance from the origin (the typical user's location), leading to  $\mathcal{R}_1<\mathcal{R}_2<\cdots$. Select the first (i.e., nearest) BS as the serving one;
    
    \State Set $S_1\leftarrow 0$; 
    
    \For{\( i_1 = 1 \) to \( N_1 \)}


        \State Sample \( \calX_{1,i_1} \) from $\calX_1 \equiv \boldsymbol{\mathcal{B}}$, where $\boldsymbol{\mathcal{B}}=(b_i)_{i>1}$ is the sequence of i.i.d. Bernoulli random elements with mean $\zeta$, indicating whether the \(i\)-th BS interferes at the typical user; for single-interferer case, retain first non-zero \(b_i\) and set others to zero;
    
        
        \State Set $S_0\leftarrow 0$; 
        
        \For{\( i_0 = 1 \) to \( N_0 \)}
        
            \State Sample \( \calX_{0,i_0} \) from \( \calX_0 \equiv \calHbold \), where $\calHbold$ is Rician fading channels between the typical user and BSs;   
            
            \State Compute \( Q_{i_0}\equiv\mathrm{SIR} \) from \eqref{eq:SIR44};
            
            \State
            {\bf if} \( Q_{i_0}
            > q \) { \bf then} \( S_0 \leftarrow S_0 +1 \);

        \EndFor
        
        \State Compute conditional probability $P_{1,i_1}^{(2)}=S_0 / N_0$;

        \State
        {\bf if } {$P_{1,i_1}^{(2)}> p_1 $} { \bf then }
              \( S_1 \leftarrow S_1 +1 \);
    \EndFor
     \State Compute conditional probability $P_{2,i_2}^{(2)}=S_1 / N_1$;

     \State
    {\bf if }{$P_{2,i_2}^{(2)}> p_2 $} {\bf then }
         \( S_2 \leftarrow S_2 +1 \);
\EndFor

\State Set \( R_{|2|}(p_1, p_2; q) = S_2 / N_2 \);

\end{algorithmic}
\end{algorithm}

To analyze the derived formulations and provide supporting discussions,
we present second-order spatial MD reliability results for both single-interferer and multi-interferer scenarios, and compare them with first-order spatial MD reliability as well as conventional (non-MD) coverage probability.  The pseudocode of the Monte Carlo simulations corresponding to the second-order MD reliability expressions \eqref{eq:thdskf3} and \eqref{eq:hfdsnxmmzevd} is presented in Algorithm 1. 
For all scenarios, we consider $t_{\mathrm{th}} = 1~\mathrm{ms}$, $l = 256$~bits, and $\plexp = 3.5$~\cite{10142008}. In Figs.~\ref{fig:can_vs_p2} and~\ref{fig:can_vs_p1}, we have illustrated the second-order spatial MD reliability $R_{[2]}$ versus the BSs' interfering target reliability $p_2$ and the link target reliability $p_1$. In Fig.~\ref{fig:can_vs_w}, we have evaluated the minimum bandwidth required to guarantee the second-order spatial MD reliability $R_{[2]}$ represented in~\eqref{eq:thdskf3}, and compare it with the bandwidth required to guarantee the first-order and zeroth-order (non-MD) reliabilities, $R_{[1]}$ and $R_{[0]}$, respectively. It can be observed from Fig.~\ref{fig:can_vs_p2} that the closed-form expression for the second-order MD reliability in~\eqref{eq:thdskf3} precisely matches the results obtained via Monte Carlo simulations for the single-interferer scenario. Furthermore, the approximate multi-interferer MD reliability representation formulated in~\eqref{eq:hfdsnxmmzevd} provides a valid and relatively tight lower bound.
The following observations can also be drawn from the numerical results:
\begin{figure}
    \centering
    \includegraphics[width=0.9\linewidth]{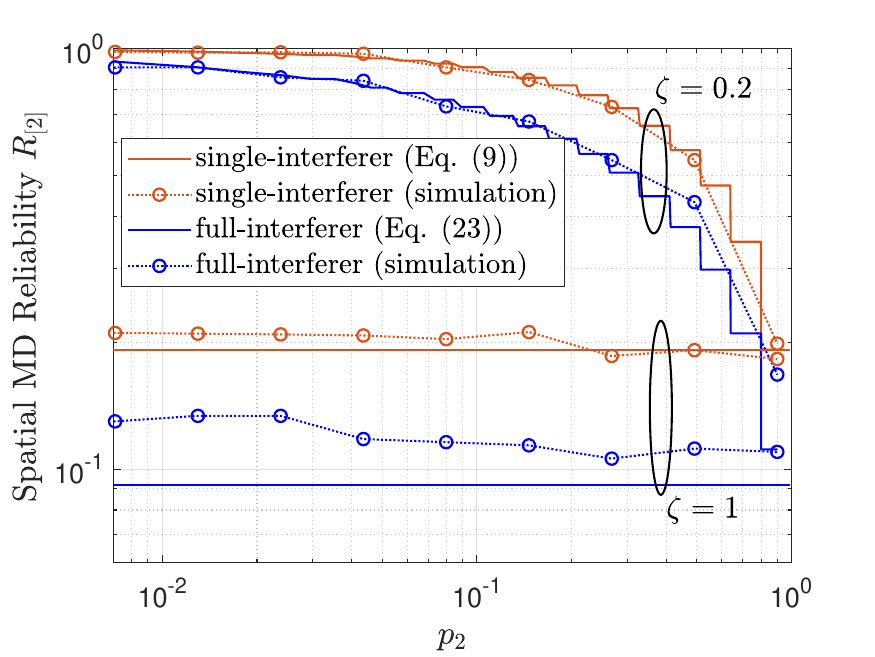}
    \caption{Single-interferer and multi-interferer spatial MD reliability $R_{[2]}$ versus interferer target reliability $p_2$, for link target reliability $p_1=0.999$, $\zeta\in\{0.2,1\}$ and $W=10$ MHz.
    }
    \label{fig:can_vs_p2}
\end{figure}
\begin{figure}
    \centering
    \includegraphics[width=0.9\linewidth]{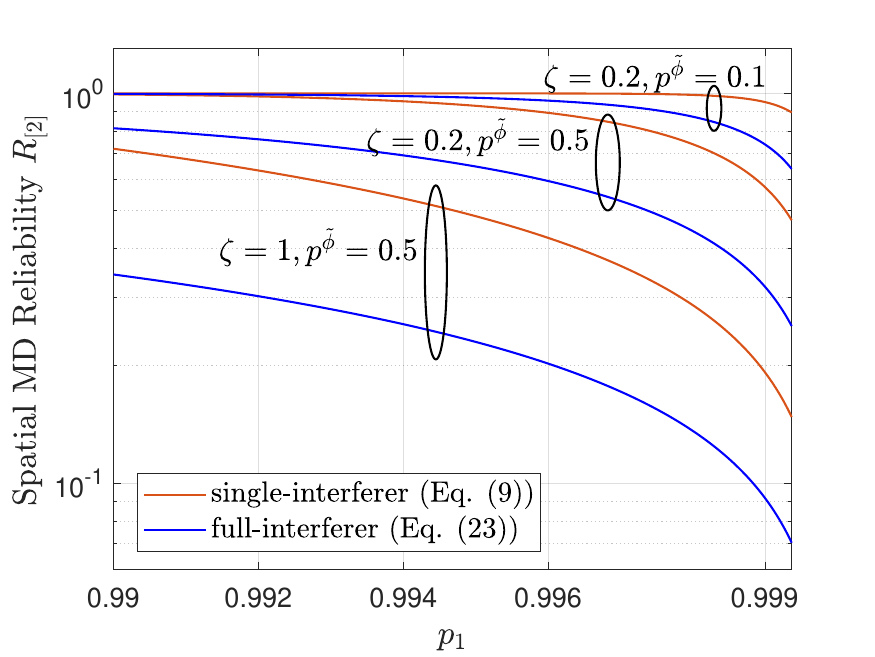}
    \caption{Single-interferer and multi-interferer spatial MD reliability $R_{[2]}$ versus  link target reliability $p_1$, for interferer target reliability $p_2\in\{0.1,0.5\}$, $\zeta\in\{0.2,1\}$ and $W=10$ MHz.
    }
    \label{fig:can_vs_p1}
\end{figure}
\begin{figure}
    \centering
    \includegraphics[width=0.9\linewidth]{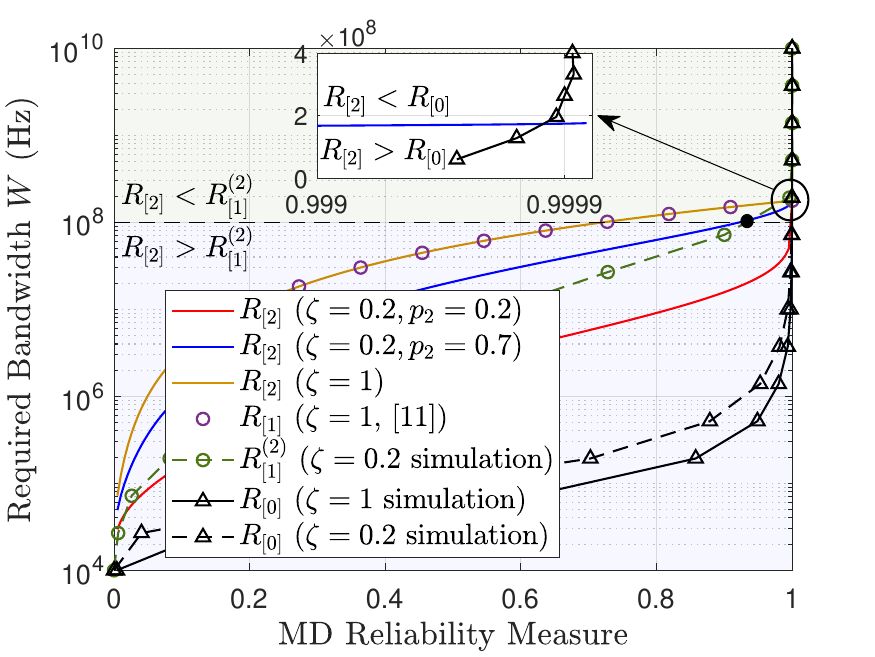}
    \caption{The bandwidth required to support MD reliability in the single-interferer scenario considering second-order MD, first-order MD, and non-MD reliability measures. 
    $\zeta\in\{0.2,1\}$, $p_1=0.999$ for $R_{[2]}$ and $R_{[1]}$, and $p_2\in\{0.2,1\}$ for $R_{[2]}$. 
    }
    \label{fig:can_vs_w}
\end{figure}
\begin{itemize}
    \item 
    As shown in Figs.~\ref{fig:can_vs_p2} and~\ref{fig:can_vs_p1}, the second-order MD reliability decreases with increasing target reliabilities $p_2$ and $p_1$. This is because a higher target value at any layer imposes stricter reliability requirements,  which is then propagated to the upper layers of the framework in Fig. \ref{fig:higher-struct}.

    \item While the MD reliability is independent of $p_2$ for $\zeta = 1$, Fig.~\ref{fig:can_vs_w} illustrates that for $\zeta < 1$, increasing $p_2$ toward unity eventually leads to a lower-bound MD reliability equal to the MD reliability measure relating to the $\zeta = 1$ case. This lower bound is seen to be approximately 0.2 and 0.09 for the single- and multi-interferer scenarios, respectively. This behavior arises because, for $\zeta = 1$, the elements in $\tilde{\Phibold}$ become fixed given $\Phibold$. 
    In this case, the middle-layer no longer influences the outcome. 
    Additionally, it is observed that smaller values of $\zeta$ lead to a higher MD reliability across all scenarios, which is attributed to the use of a thinner PPP to represent the interfering BSs.

    \item 
    

    Noting that higher-order reliability measures enforce nuanced target success probabilities at multiple sub-layers of the reliability analysis, it is important to investigate the bandwidth required to satisfy these measures and to compare the bandwidth demands of higher-order and lower-order MD reliabilities. This comparison is presented in Fig.~\ref{fig:can_vs_w}, which shows  the bandwidth required to provide $R_{[2]}^{(2)}\equiv R_{[2]}$, $R_{[1]}^{(2)}$, and $R_{[0]}$ measures. For the first-order MD, we have lumped the random variables $\calX_1=\boldsymbol{\mathcal{B}}$ and $\calX_2=\Phibold$ together to compute $R_{[1]}^{(2)}=\mathbb{P}(P_1^{(2)}>p_1)$. However, when $\zeta=1$, the middle layer corresponding to $\boldsymbol{\mathcal{B}}$ is effectively bypassed, resulting in $R_{[1]}^{(2)}=R_{[1]}^{(1)}\equiv R_{[1]}$. A closed-form solution to this simple case has been previously obtained in~\cite{10142008} for the bandwidth $W$ corresponding to $R_{[1]}$. It is seen that the curves corresponding to $R_{[2]}$ and $R_{[1]}$ for $\zeta = 1$ (the third and fourth curves in the plot legend) exactly coincide, which are completely in agreement with our derived expression for the first-order MD reliability in this special case. 
  Furthermore, while all reliability measures improve with increasing bandwidth \( W \), the \textit{rate of improvement} is consistently greater for higher-order MD reliabilities at lower values of \( R_{[2]} \). Specifically, there exist threshold values of \( W \) beyond which lower-order reliabilities surpass higher-order ones, and below which the opposite holds. For example, in the case of \( \zeta = 0.2 \), the second-order MD reliability \( R_{[2]} \) corresponding to $p_2=0.7$ exceeds the first-order reliability \( R_{[1]}^{(2)} \) when \( W < 10^8 \) Hz, and similarly, we have \( R_{[2]} > R_{[0]} \) when \( W < 1.8 \times 10^8 \) Hz. A comparison across all plots reveals that for moderate values of \( R_{[2]} \) not very close to unity, achieving higher-order reliabilities generally requires more bandwidth than their lower-order MD and especially non-MD counterparts.

\end{itemize}

\subsection{Spatial-Spectral-Temporal MD Reliability in UWB THz Communication}
\label{sec:case_study_2}
In what follows, we apply the higher-order MD reliability analysis to an ultra-wideband (UWB) THz network leveraging pseudorandom slowly time-varying frequency hopping carrier assignment \cite{monemi2025higherconference}. 
\subsubsection{System Model}
    The statistics of the carrier frequency can influence the overall reliability of a communication link. Incorporating the spectral domain in the reliability analysis is more important when dealing with UWB communications. For example, consider a UWB communication through frequency hopping spread spectrum (FHSS) where carriers assigned to users  vary over time according to a pseudorandom policy, spanning the entire available spectrum. This provides benefits such as security and robustness making the communication more resilient against interference and jamming. While the impact of frequency might be negligible in the reliability measure in applications requiring a low amount of spectrum, this is not the case for UWB applications. 


\begin{table}
    \centering
    \caption{Parameters for Numerical Results in Section \ref{sec:case_study_2}}
    \begin{tabular}{|l|l|l|}
        \hline
        \textbf{Parameter}                     & \textbf{Description} 
        & \textbf{Value}
        \\ \hline
        \rule{0pt}{8pt}
        $(\underline{f}^{(2)}_{\ },\overline{f})$ 
        & Frequency range in Scenario 1 
        & $(340,375)$ GHz 
        \\ \hline
        \rule{0pt}{8pt}
        $(\underline{f}^{(1)}_{\ },\overline{f})$ 
        & Frequency range in Scenario 2 
        & $(325,375)$ GHz 
        \\ \hline
        $\lambda$ 
        & Intensity of the PPP
        & $1.5\times 10^{-3} \mathrm{\frac{1}{m^{2}}}$
        \\
        \hline
        $(G_{\mathrm{T}},G_{\mathrm{R}})$ 
        & Transmit and receive antenna gains
        & $(25,25)$ dB
        \\			
        \hline
        $W$
        & Bandwidth
        & 1 GHz
        \\			
        \hline
        $l$
        & No. of bits to be received in time  $t_{\mathrm{th}}$
        & $1000$
        \\			
        \hline
        $t_{\mathrm{th}}$
        &User-plane deadline threshold 
        & $10\ \mu$s
        \\			
        \hline
        $K$ 
        & Rician shape factor
        & $2$  
        \\			
        \hline
        $P_{\mathrm{T}}$ 
        & Transmit power
        & 0.1 W 
        \\		
        \hline
        $k(f)$
        & Molecular absorption coefficient
        & See Fig. 1 in \cite{8568124}
         \\		
        \hline
    \end{tabular}
    \label{tbl:simulation_params}
\end{table}

Consider an FHSS UWB network of randomly located nodes communicating in THz band where each user is assigned a carrier frequency, selected from a pseudorandom sequence generated for that user. While the sequence generation process is deterministic, the resulting frequency hopping pattern is stochastic to an external observer, where the corresponding pdf is determined by the carrier assignment
algorithm. 
Here the statistics of the varying carrier frequency can highly affect the reliability. This is because the large-scale path loss is a function of the frequency, especially at THz bands where the molecular absorption is a frequency-dependent factor that highly affects the signal attenuation. Following Example 1, considering that co-channel interference is negligible 
and all links follow same channel fading statistics,
we can express the MD reliability according to \eqref{eq:Rncompact}, where $Q=1/t_l(\mathrm{SNR}(\calH,\calF,\R)$, in which the ordered collections of random variable are $\calX_0\equiv\calH$, $\calX_1\equiv\calF$ and $\calX_2\equiv\R$. These are scalars corresponding to the small-scale fading, carrier frequency, and the distance between some user in the network and its nearest BS. For each user, the carrier frequency is a pseudorandom variable selected according to some pdf determined by the carrier assignment algorithm.  
Similar to Example 1 and considering the line-of-sight (LoS) THz channel model 
\cite{8568124}
as well as the simple case of orthogonal carrier allocation, we can formulate the SNR as 
\begin{align}
    \label{eq:tlTHz}
    \mathrm{SNR}=\frac{P_{\mathrm{T}}  G_{\mathrm{T}} G_{\mathrm{R}}c^2}{(4\pi f)^2} \left( \frac{ h  r^{-2} e^{{-k(f) r}}}{N_0 W}\right),
\end{align}
where $r$ is the distance, $k(f)$ is the molecular absorption coefficient at frequency $f$, and $h$ is the the small-scale fading coefficient. We aim to calculate the MD reliability of delivering $l$ bits with a time delay lower than a user-plane deadline threshold $t_{\mathrm{th}}$, given the target temporal reliability $p_1$ and target spectral reliability $p_2$.


\begin{figure*}
    \centering
    \includegraphics[width=0.95\linewidth]{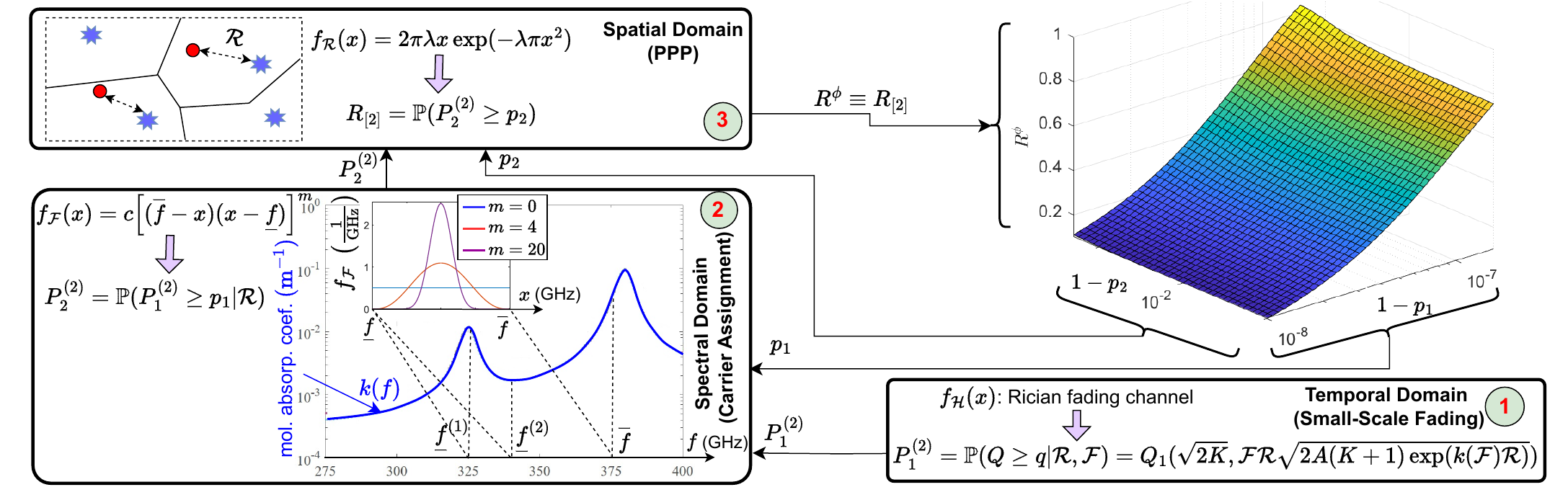}
    \caption{Calculation of the second-order spatial MD reliability $R^{\mathrm{\phi}} \equiv R_{[2]}$ versus $q$, $p_1$ and $p_2$.}
    \label{fig:sim3D}
\end{figure*}
We consider a THz network wherein BSs are scattered according to PPP with density $\lambda$, 
and each user is assigned to the nearest BS. Accordingly, the pdf of the distance is $f_\R(r)=2\pi\lambda r\mathrm{exp}(-\lambda \pi r^2)$. We have adopted the molecular absorption coefficient according to Fig. 1 in \cite{8568124} for the frequency range from $\underline{f}=275$ GHz to $\overline{f}=325$ GHz, where the corresponding coefficient $k(f)$ is depicted in Fig. \ref{fig:sim3D}. We assume that each user is assigned a carrier frequency at each time step where the carrier is selected according to some pdf $f_\F$ supported on $[\underline{f},\overline{f}]$. The pseudorandom carrier assignment is commonly considered to have uniform distribution $U(\underline{f},\overline{f})$ to allow effective spreading of the signal across the available bandwidth. 
To investigate the impact of frequency domain pseudorandom carrier assignment in the overall MD reliability measure, we adopt the more general model
\begin{align}
\label{eq:fF}
f_\F(x)=c\left[(\overline{f}-x)
(x-\underline{f})\right]^m, 
\end{align}
where $m$ is the shape factor and $c=(\overline{f}-\underline{f})^{-1-2m}/\beta(1+m,1+m)$ in which $\beta$ is the beta function. As seen in Fig. \ref{fig:sim3D}, adjusting the shape factor $m$ results in different pdf models. For $m=0$, it is the uniform distribution, and as $m \rightarrow \infty$, it approaches the Dirac delta function at $(\underline{f}+\overline{f})/2$.

Finally in the temporal domain, noting that THz communication is mostly achieved in LoS for short ranges, a Rician fading channel model with pdf $f_\HH$ having shape factor $K$ is assumed. For the sake of simplicity, we are not including a blockage model for communication between the BS and user, as considered in some works in the literature \cite{10643015,10063854}. 

\subsubsection{Calculation of the second-order spatial MD reliability}\ We aim to calculate the second-order spatial MD reliability $R_{[2]}$ for the stated THz network. 
Fig. \ref{fig:sim3D} illustrates the steps taken according to  \eqref{eq:MDitterative} to calculate the MD reliability. In the first step, we formulate $P_1^{(2)}=
\mathbb{P}(Q> q\mid
\R,\F)$ by taking the expectation with respect to fading random elements in the temporal domain. Considering the Rician fading channel model and the representation of $t_l$ for THz channels expressed in \eqref{eq:tlTHz}, after some mathematical manipulations (see {\ref{sec:apxIA1}}), $P_1^{(2)}$ can   obtained as 
\begin{align}
\label{eq:P1case}
    {P}_1=Q_1(\sqrt{2K},\F \R\sqrt{2c_1q(K+1)\exp(k(\F)\R)
}),
\end{align} 
where $c_1=\frac{ N_0W(4\pi)^2}{P_{\mathrm{T}}  G_{\mathrm{T}} G_{\mathrm{R}}c^2}   $ and $q=(2^{\frac{l}{W t_{\mathrm{th}}} } -1)$, and $Q_1$ is the first-order Marcum Q-function. In the second step, given $P_1^{(2)}$ and the target temporal-domain link reliability $p_1$, and considering the adopted models for molecular absorption coefficient $k(f)$ as well as the pdf for carrier assignment $f_\F$, we can formulate $ P_2^{(2)}=
\mathbb{P}(P_1^{(2)}> p_1\mid\R)$. Finally, considering the pdf of $f_\R$ obtained from PPP where $\R$ is the distance between the user and the nearest BS, in the third step we can obtain $R_{[2]}\equiv R^{\mathrm{\phi}}$ corresponding to a target spectral reliability  $p_2$ by solving $R_{[2]}=
\mathbb{P}(P_2^{(2)}> p_2)$. 
To provide the analytical solution, we have considered two scenarios. In Scenario 1, corresponding to Figs. \ref{fig:sim_vs_Rh} and \ref{fig:sim_vs_alpha}, we explore the MD reliability analysis for a fixed bandwidth of ${\rm BW}=\overline{f}-\underline{f}^{(2)}$ corresponding to a {\it monotonically increasing} part of $k(f)$ in the frequency range $(\underline{f}^{(2)},\overline{f})$. In Scenario 2, we investigate the MD reliability analysis for a variable frequency range of $(\underline{f}^{(1)},\underline{f}^{(1)}+{\rm BW})$, where ${\rm BW}\in[0,\overline{f}-\underline{f}^{(1)}]$ lies within a more general {\it non-monotonic} part of $k(f)$. The corresponding values considered for $\underline{f}^{(1)}$, $\underline{f}^{(2)}$ and $\overline{f}$ are shown in Fig. \ref{fig:sim3D} and Table \ref{tbl:simulation_params}. An analytical closed-form solution for Scenario 1 and a low-complexity numerical solution scheme for Scenario 2 has been presented in {\bf \ref{sec:apxIA}} and {\bf \ref{sec:apxIB}} respectively.

\subsubsection{Numerical Results and Discussion}  
\label{sec:numerical_casestudy_2}
Fig.~\ref{fig:sim3D} presents a 3D MD reliability diagram for \(R^{\mathrm{\phi}}\) as a function of \(p_1\) and \(p_2\). Additionally, 2D plots of MD reliability versus \(p_1\), \(p_2\), and bandwidth (BW) under various simulation parameters are shown in Figs.~\ref{fig:sim_vs_Rh}, \ref{fig:sim_vs_alpha}, and \ref{fig:sim_vs_BW}, respectively. 
\begin{figure}
    \centering
\includegraphics[width=0.9\linewidth]{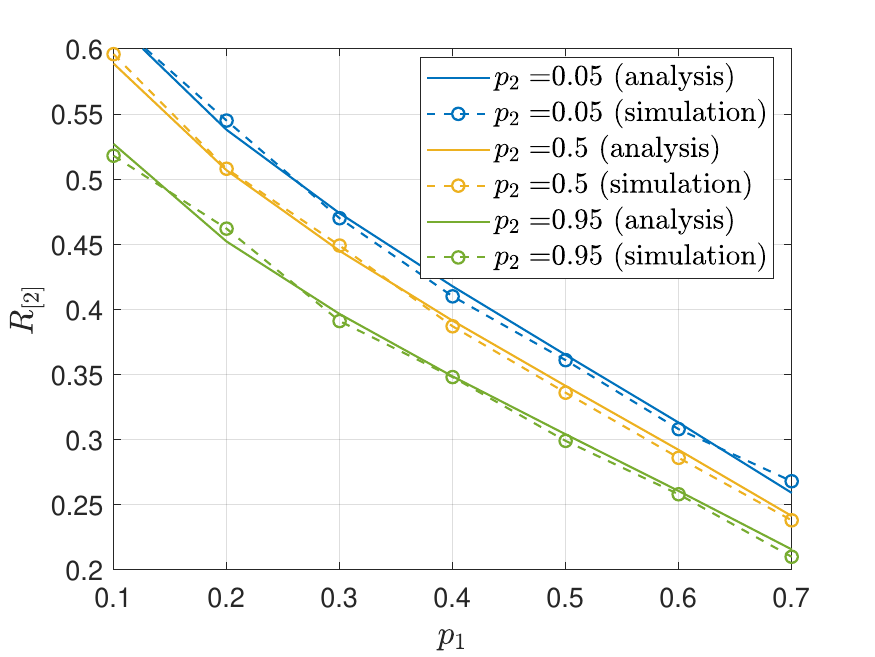}
    \caption{Monte Carlo simulations corresponding to the analytical results versus $p_1$ and $p_2$, using normalized QoS threshold $q=1$, spectral shaping factor $m=1$, $c_1=0.01/\overline{f}^2$ as defined in \eqref{eq:739re52}, and $(\underline{f},\overline{f})=(340,375)$ GHz.
    }
    \label{fig:monte_carlo_verify}
\end{figure}
  Before analyzing these figures, we supported the derived analytical results in Fig.~\ref{fig:monte_carlo_verify} with Monte Carlo simulations. Note that MD-based Monte Carlo simulations are generally more computationally demanding than conventional approaches, even for a two-level, first-order MD evaluation, due to the nested conditional loops. This complexity increases significantly in second-order, three-layer MD computations, particularly when the per-layer success probabilities approach zero or one. In our numerical results in Figs.~\ref{fig:sim_vs_Rh}, \ref{fig:sim_vs_alpha}, and \ref{fig:sim_vs_BW}, the target threshold \(p_1\) is set extremely close to unity (e.g., 0.99999 or higher), representing the very high link reliability required for URLLC services~\cite{10142008}. Under such conditions, Monte Carlo simulations within the three-layer MD framework become computationally highly resource-intensive. To make the MD-based Monte Carlo simulations more tractable, we used intermediate parameter values, including normalized threshold \(q\) as well as considering \(p_1\) and \(p_2\) not too close to zero or one. Fig.~\ref{fig:monte_carlo_verify} confirms the validity of our analytical formulations. Since the derived expressions are independent of specific parameter values, we proceed with the analysis across various scenarios in Figs.~\ref{fig:sim_vs_Rh}, \ref{fig:sim_vs_alpha}, and \ref{fig:sim_vs_BW}, using the practical parameter values listed in Table~\ref{tbl:simulation_params}.

Key observations from the numerical results include the following:

\begin{itemize}
    \item First, it is seen how the spatial MD reliability measure is a monotonically decreasing function of both temporal and spectral reliability measures. For example, it is observed in Fig. \ref{fig:sim_vs_Rh} that for $m=0$ and $p_1=1-7\times 10^{-8}$, increasing $p_2$ from 0.9 to 0.99 decreases the spatial MD reliability $R^{\mathrm{\phi}}$ from 0.8 to 0.74. The monotonically decreasing property is justified by noting that guaranteeing higher reliability measures in the temporal and spectral domains is achievable in a smaller portion of the network area, corresponding to a smaller spatial MD reliability.
    
    \begin{figure}
    \centering
    \includegraphics[width=0.9\linewidth]{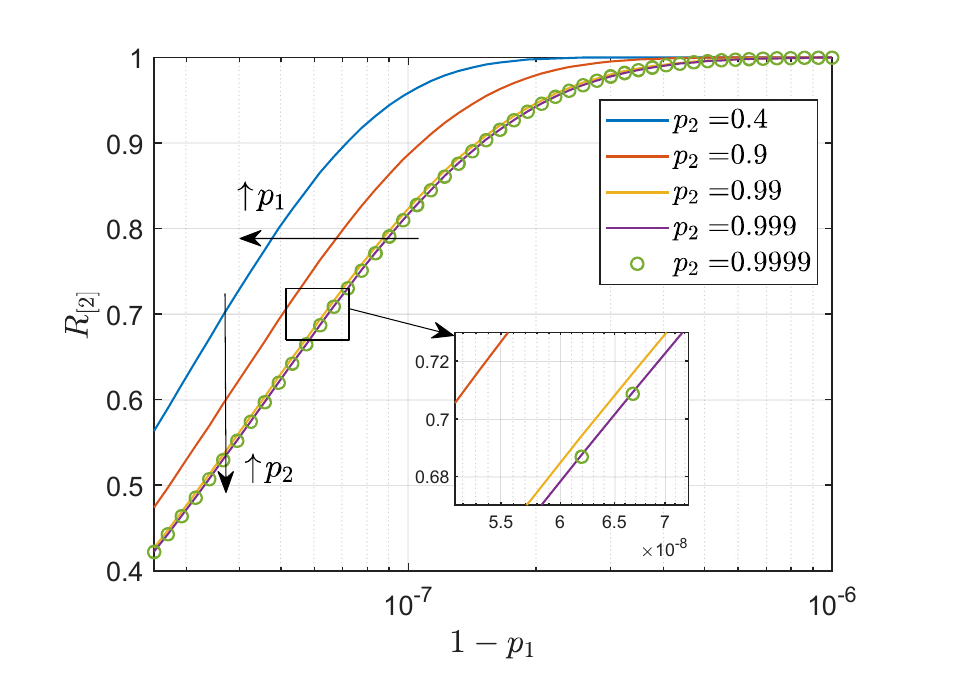}
    \caption{Spatial MD reliability $R_{[2]}$ versus temporal ($p_1$) and spectral ($p_2$) target reliabilities for $m=0$.}
    \label{fig:sim_vs_Rh}
\end{figure}
\begin{figure}
    \centering
    \includegraphics[width=0.9\linewidth]{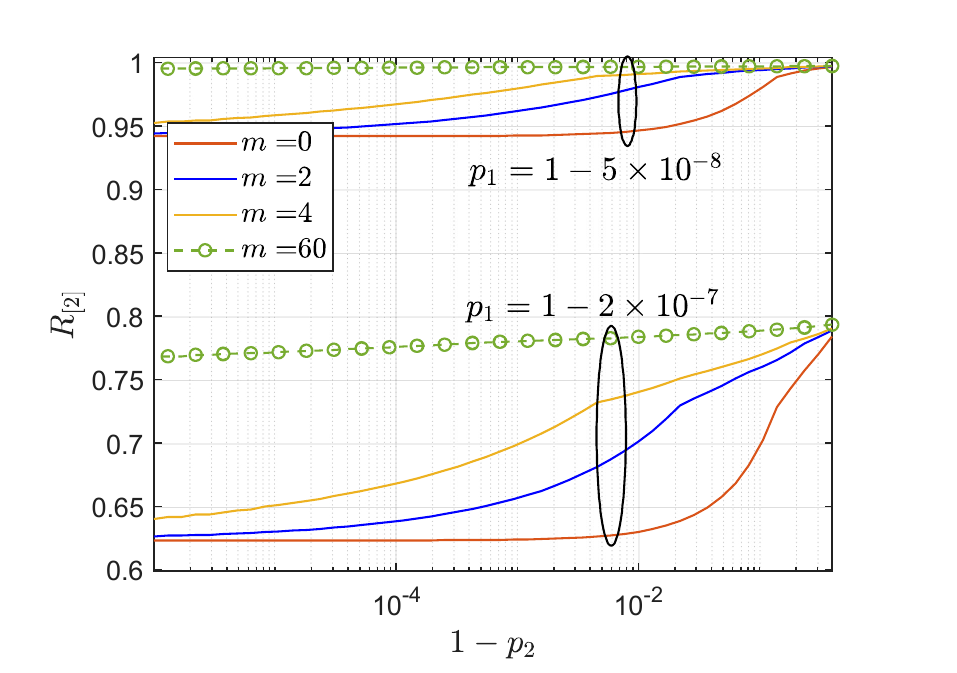}
    \caption{Spatial MD reliability $R_{[2]}$ versus temporal ($p_1$) and spectral ($p_2$) target reliabilities for different values of $m$.}
    \label{fig:sim_vs_alpha}
\end{figure}

    \item The higher-order MD reliability analysis can give insights into the impact level of the target reliability of each dimension on the overall MD reliability measure. For example, as seen in Fig. \ref{fig:sim_vs_alpha}, going toward higher values of the spectral pdf shape factor (e.g.,  $m=60$) increases the MD reliability at the cost of not effectively spreading the signal over the whole spectrum, leading to lower resiliency and higher risk of jamming.
    

    \item Fig. \ref{fig:sim_vs_BW} shows another feature of the MD reliability of wideband THz communications. For any given spectral target reliability $p_2$ in Scenario 2 wherein $k(f)$ is not a monotonically increasing function in the available frequency range $(\underline{f}^{(1)},\overline{f})$, the overall MD reliability is potentially optimal at some certain bandwidth value shown as filled circles, below and after which the MD reliability measure is smaller. The reason behind this relates to the mathematical formulation of $P_2^{(2)}$ presented in \eqref{eq:83236455} in {\ref{sec:apxIB2}}. It is seen that given $\R$, for low values of the bandwidth,  $\F\R\sqrt{\exp(k(\F)\R)}$ in \eqref{eq:83236455} is potentially decreasing in terms of $\F$. This is because the non-linearly decreasing term $\sqrt{\mathrm{exp}(k(\F)\R)}$ for frequency range close to $\underline{f}^{(1)}$ is potentially the dominant term compared to the linearly increasing term $\F\R$, leading to this function be finally decreasing in terms of $\F$ for low bandwidth values. This increases the probability of $P_2^{(2)}$ in \eqref{eq:83236455}, leading to a higher MD reliability. However, as the bandwidth increases, $\F\R\sqrt{\mathrm{exp}(k(\F)\R)}$ becomes an increasing function of $\F$ after some point $\F^*\leq \mathrm{argmin}_{f \in(\underline{f}_1^{(1)},\overline{f})}
    \{k(f)\}$ since finally both exponential and linear terms will be monotonically increasing for frequencies higher than $\mathrm{argmin}_{f \in(\underline{f}_1^{(1)},\overline{f})}
    \{k(f)\}$, leading to lower success probability in \eqref{eq:83236455} at such frequencies compared to that in $\F^*$, as shown in Fig. \ref{fig:sim_vs_BW}.
\end{itemize}

\begin{figure}
    \centering
    \includegraphics[width=0.9\linewidth]{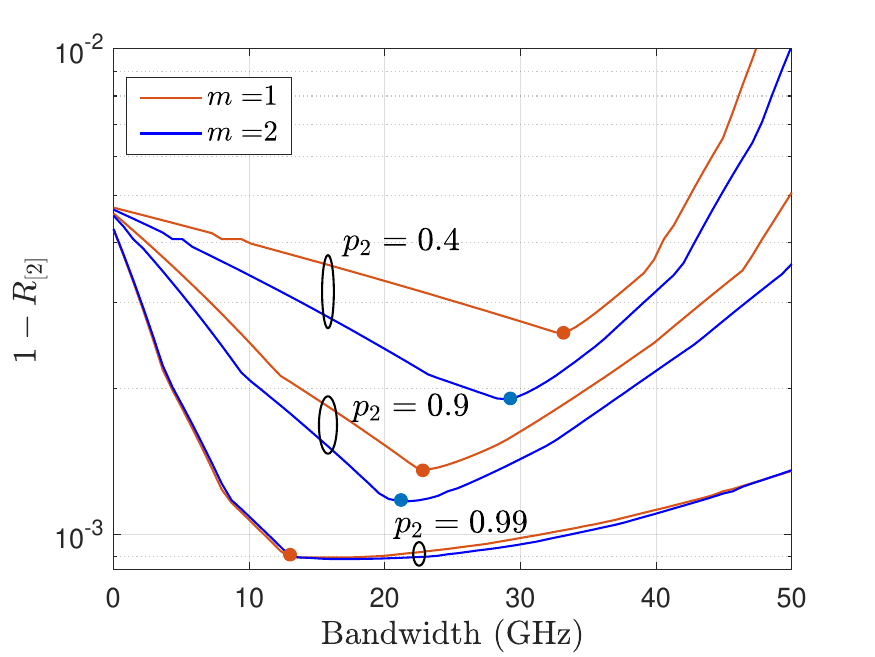}
    \caption{Spatial MD reliability $R_{[2]}$ versus bandwidth for different values of $m$ and spectral target reliability $p_2$.}
    \label{fig:sim_vs_BW}
\end{figure}

\section{Conclusions}
In this paper, we extended the meta distribution (MD) reliability analysis beyond conventional first-order spatiotemporal schemes. By structuring MD reliability in a hierarchical framework, we introduced a mathematical representation for higher-order MD reliability characterization, where the overall MD reliability is formulated in terms of the desired QoS and target reliability thresholds in multiple domains.
More specifically, we proposed a framework for the analysis of higher-order MD reliability of wireless networks considering three levels of temporal dynamicity of random elements including fast, slow and static random elements, where the MD at each layer is leveraged to be used in calculating the MD of the higher layer. 
Furthermore, we investigated the analysis of the second-order MD reliability for two applications in wireless networks that can take advantage of the characterized higher-order MD reliability analysis approach.
The first involved the second-order MD reliability for a canonical stochastic network setup wherein the interfering BSs correspond to slowly time-varying Bernoulli random variables. The second explored a second-order spatial-spectral-temporal MD reliability for an ultra-wideband (UWB) frequency-hopping spread spectrum (FHSS) THz network. 
For both applications, we provided detailed analytical derivations and numerical results. Our analysis revealed how target reliabilities in each domain influence the overall spatial MD reliability, providing nuanced insights into system performance that go beyond the capabilities of non-MD or first-order MD reliability analyses.


\renewcommand{\thesection}{Appendix I} 
\section{\ }
\subsection{Proof of Theorem \ref{th:1}}
\label{sec:proof_th1}

From \eqref{eq:MDitterative} and \eqref{eq:MDitterative_rel}, we need to calculate $P_1^{(2)}$, $P_2^{(2)}$ and $R_{[2]}$ in a hierarchical scheme, represented as follows:
\begin{subequations}
\begin{align}
    \label{eq:P1cc}
     P_1^{(2)}&=  \mathbb{P}
     (
        \mathrm{SIR} >q \mid \Phibold,\boldsymbol{\mathcal{B}}
    )
    \\
    \label{eq:P1cc2}
     P_2^{(2)} &=\mathbb{P}(P_1^{(2)}> p_1\mid \Phibold)
     \\
     \label{eq:P1cc3}
     R_{[2]} &=\mathbb{P}(P_2^{(2)}> p_2).
\end{align}
\end{subequations}
We first calculate $P_1^{(2)}$ from \eqref{eq:P1cc} as follows: 
\begin{align}
    P_1^{(2)}&= \mathbb{P}\left(\frac{\calH_1 \R_1^{-\plexp}}{\calHtil_1 \Rtil_1^{-\plexp}} >q \bigg| \Phibold, \boldsymbol{\mathcal{B}} \right) 
    \notag \\
    &= \mathbb{P}\left( \frac{\calH_1}{\calHtil_1} > q  \left( \frac{\R_1}{\Rtil_1} \right)^{\plexp} \Bigg| \Phibold,\boldsymbol{\mathcal{B}} \right)
    \notag \\
\label{eq:heirizznf88462}
    &
    {=} \left[{1 + q \left( \frac{\R_1}{\Rtil_1} \right)^{\plexp}}\right]^{-1},
\end{align}
where the last equality follows from the fact that ${F}_{\calH_i/\calH_j}(x)=\frac{x}{1+x},\forall i\neq j$ \cite{9389789}.  The random variable $P_2^{(2)}$ is then formulated by calculating the success probability for $P_1^{(2)}>p_1$ expected over the random elements $\boldsymbol{\mathcal{B}}$. This is represented as follows:
\begin{align}
    P_2^{(2)} &=\mathbb{P}(P_1^{(2)}> p_1\mid \Phibold)
\notag \\
    &= 
    \mathbb{P} \left(
    \left[1 + q  \left( {\R_1}/{\tilde{\R}_1} \right)^{\plexp}\right]^{-1} >p_1\bigg| \Phibold \right)
 \notag \\
      & \overset{(a)}{=}
    \mathbb{P}\left(
    \left({\R_1}/{\Rtil_1}\right)^{\plexp} <
    \left({p_1^{-1}-1}\right){ q^{-1}}  \bigg| \Phibold \right)
\notag \\
    & \overset{(b)}{=}
    \mathbb{P}\left(
    \Rtil_1>\hat{p}_1 \R_1 \mid \Phibold
    \right),
\notag \\
 & \overset{(c)}{=}
    \sum_{i=2}^{\infty}\mathbb{P}\left(
    \R_i>\hat{p}_1 \R_1 \mid \Phibold, \tilde{\phi}_i
    \right)\mathbb{P}(\tilde{\phi}_i)
\notag \\
    &\overset{(d)}{=}\sum_{i=2}^{\infty}\mathbb{P}\left(
    \R_i>\hat{p}_1 \R_1 \mid \Phibold
    \right)\times
    \underbrace{\zeta (1-\zeta)^{i-2}}_{c_i} 
\notag \\
\label{eq:hjsd6835}
    &\overset{(e)}{=}\sum_{i=2}^{\infty} c_i\underbrace{\mathbf{1}\left(
    \R_i>\hat{p}_1 \R_1
    \right)}_{X_i} .
\end{align}
The derivation of (b) follows directly from (a), where \( \hat{p}_1 \) is given in \eqref{eq:p1hat}. Let \( \tilde{\phi}_i \) denote the event that the \( i \)-th nearest BS is the (first) interferer. In what follows, we consider the case where \( \hat{p}_1 > 1 \); otherwise, \( P_2^{(2)} \), and consequently the overall MD reliability, is trivially found to be equal to one.
 Expression (b) can be rewritten as (c) by marginalizing over the events $\tilde{\phi}_i,\forall i$. We observe that if the first interferer corresponds to the 
$i$-th BS, this requires the $i$-th BS to be interfering (with probability $\zeta$) and all BSs with indices \( 2 \leq i' \leq i-1 \) to be non-interfering, each with probability \( 1 - \zeta \). Therefore, we have \( \mathbb{P}(\tilde{\phi}_i) = \zeta (1 - \zeta)^{i - 2} \), as represented in (d).
 Finally, noting that conditioned on \( \Phibold \), the values of \( \mathcal{R}_i,\forall i \) are deterministic, the probability \( \mathbb{P} \) in (d) reduces to the indicator function $\mathbf{1}$ in (e).

Let \( N(r_i \mid r_1) \) denote the number of BSs within the ball \( \{ r \leq r_i \} \), given that the first BS is located at distance \( r_1 \). Observe that for any \( i \geq 2 \), the event \( \{ \mathcal{R}_i > \hat{p}_1 \mathcal{R}_1 \} \) is equivalent to \( \{ N(\hat{p}_1 \mathcal{R}_1 \mid \mathcal{R}_1) < i \} \). 
Let \( N' \) denote the number of BSs other than the one at \( \mathcal{R}_1 \) that lie within the region \( \{ r \leq \hat{p}_1 \mathcal{R}_1 \} \). Then, we have
\[
N' = N(\hat{p}_1 \mathcal{R}_1 \mid \mathcal{R}_1) - 1.
\]
This leads to the following:
\begin{multline}
    X_i=\mathbf{1}\{ R_i> \hat{p}_1\R_1\}= \mathbf{1}\{ N(\hat{p}_1 \R_1|\R_1)<i\}
    \\
    \label{eq:yuewi83gfg}
    = \mathbf{1}(N'\leq i-2), 
\end{multline}
which results in
\begin{multline}
    P_2^{(2)}
    =
    \sum_{i=2}^{\infty} c_i \mathbf{1}(N'\leq i-2) 
   = \sum_{i=N'+2}^{\infty} c_i=\sum_{i=N'}^{\infty}  \zeta(1-\zeta)^i
   \\
   \label{eq:yuewi83gfg1}
   = (1-\zeta)^{N'}.
\end{multline}

Finally, the second-order MD reliability is expressed as
\begin{multline}
     R_{[2]}=  \mathbb{P}(P_2^{(2)}>p_2)
     =\mathbb{P}\left(  (1-\zeta)^{N'}>p_2\right)
     \\
\label{eq:djg0374zz}
     =\mathbb{P}\left(
     N'<\frac{\ln (p_2)}{\ln(1-\zeta)}
     \right) =
    \mathbb{P}\left(
     N'\leq\left\lfloor \frac{\ln (p_2)}{\ln(1-\zeta)}\right\rfloor
     \right)
\end{multline}
To evaluate \eqref{eq:djg0374zz}, we require the distribution of \( N' \), i.e., \( \mathbb{P}(N' = n) \). First we investigate the case of $n=0$. Given that the first (serving) BS is located at a distance $\R_1$, $\mathbb{P}(N' = 0)$ corresponds to no additional BS being located within the distance \( \hat{p}_1 \mathcal{R}_1 \), which yields $\mathbb{P}(N' = 0) = \mathbb{P}(\mathcal{R}_2 > \hat{p}_1 \mathcal{R}_1) = \frac{1}{\hat{p}_1^2}$, 
    where the last equality follows from the fact that \( F_{\mathcal{R}_1 / \mathcal{R}_2}(x) = x^2 \) for a homogeneous PPP in \( \mathbb{R}^2 \) \cite{9072358}.
    For $n=1$, \( \mathbb{P}(N' = 1)\) corresponds to exactly one additional (non-serving) BS lying within the region \( \{ r \leq \hat{p}_1 \mathcal{R}_1 \} \), i.e., having \( \mathcal{R}_3 > \hat{p}_1 \mathcal{R}_1 \) but \( \mathcal{R}_2 \leq \hat{p}_1 \mathcal{R}_1 \). Noting the memoryless property of the PPP, this equals to $\mathbb{P}(\mathcal{R}_2 \leq \hat{p}_1 \mathcal{R}_1 < \mathcal{R}_3) = \left(1 - \frac{1}{\hat{p}_1^2}\right)\frac{1}{\hat{p}_1^2}$.
    Finally, for the general case, \( \mathbb{P}(N' = n) \) corresponds to exactly \( n \) non-serving BSs lying within \( \hat{p}_1 \mathcal{R}_1 \), i.e., the first \( n \) non-serving BSs are inside the region, while the \( (n+1) \)-th one lies outside. This results in
    \begin{align}
    \label{eq:hhfjgs64782}
        \mathbb{P}(N' = n) = \left(1 - \frac{1}{\hat{p}_1^2}\right)^n \frac{1}{\hat{p}_1^2}.
    \end{align}
From \eqref{eq:djg0374zz} and \eqref{eq:hhfjgs64782}, we conclude
\begin{align}
         R_{[2]}&=  \sum_{n=0}^{\left\lfloor \frac{\ln (p_2)}{\ln(1-\zeta)}\right\rfloor} \mathbb{P}(N'=n)
        =1-\left( 1-
        \frac{1}{\hat{p}_1^2}
        \right)^{\left\lfloor \frac{\ln (p_2)}{\ln(1-\zeta)}\right\rfloor+1 }.
\end{align}
This completes the proof.



\subsection{Proof of Lemma \ref{lm:fgsk46}}
\label{sec:proof_lm1}
\begin{figure}
        \centering
        \includegraphics[width=0.9\linewidth]{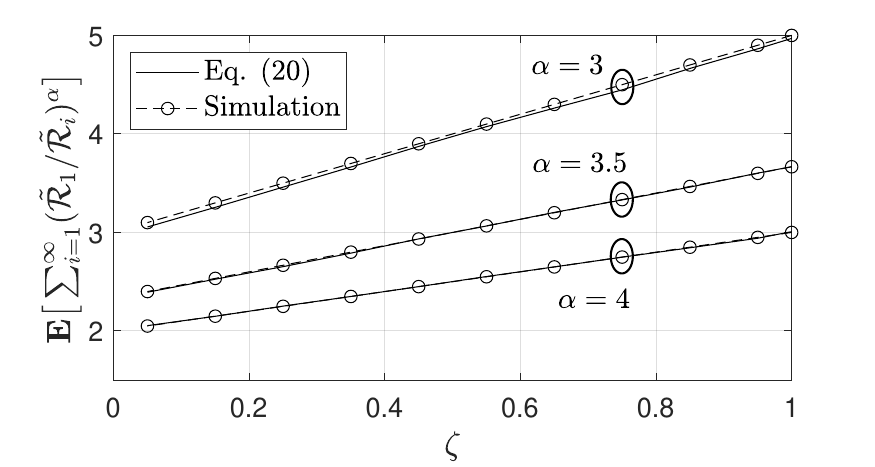}        \caption{Validation of the closed-form approximation $\mathbb{E}\left[\sum_{i=1}^{\infty}({ \Rtil_1}/{\Rtil_i})^{\plexp}\right]\approx\frac{1+\delta \zeta}{1-\delta}$ with Monte Carlo simulation.}     \label{fig:relation_vs_alpha_zeta}
    \end{figure}

We first consider the extreme cases corresponding to $\zeta=1$ and $\zeta\rightarrow 0$. For $\zeta=1$, we have $\Rtil_i=\R_{i+1},\forall i$, and thus $\mathbb{E}\big[\sum_{i=1}^{\infty}({ \Rtil_1}/{\Rtil_i})^{\plexp}\big]=\mathbb{E}\big[\sum_{i=2}^{\infty}({ \R_2}/{\R_i})^{\plexp}\big]=\frac{\plexp+2}{\plexp-2}$ \cite{10142008}. Observe that $\Rtil_i$ is a random variable which depends on $\R_1$. However, for $\zeta \rightarrow 0$, we have $\Rtil_1 \gg \R_1$, and therefore $\Rtil_i$, here denoted by $\R'_i$, can be considered independent of $\R_1$.  
    The result is then simplified to $\mathbb{E}\big[\sum_{i=1}^{\infty}({ \Rtil_1}/{\Rtil_i})^{\plexp}\big]=\mathbb{E}\big[\sum_{i=1}^{\infty}({ \R'_1}/{\R'_i})^{\plexp}\big]=1+\frac{2}{\plexp-2}=\frac{\plexp}{\plexp-2}$ \cite{6897962}. Using a linear interpolator that satisfies the boundary conditions, we obtain
    \begin{align}
        \zeta \frac{\plexp+2}{\plexp-2} + (1-\zeta) \frac{\plexp}{\plexp-2}
        =
        \frac{1+\delta\zeta}{1-\delta}.
    \end{align}
    Fig. \ref{fig:relation_vs_alpha_zeta} reveals that the derived relation tightly matches the result obtained from the Monte Carlo simulation.
    
\subsection{Proof of Theorem \ref{th:2}}
\label{sec:proof_th2}
Similar to \eqref{eq:heirizznf88462}, for the multi-interferer scenario $P_1^{(2)}$ is formulated as 
\begin{align}
    P_1^{(2)}
    &=
    \mathbb{P}\left( \frac{\calH_1 \R_1^{-\plexp}}{\sum_{i=1}^{\infty} \calHtil_i \Rtil_i^{-\plexp}} > q  \Bigg| \Phibold, \boldsymbol{\mathcal{B}} \right)
    \notag \\
\label{eq:hdjkdsiu883mmsq}
    &=
    \mathbb{P}\left( \frac{\calH_1 }{\calHtil_1} >  q \left(\frac{\R_1}{\Rtil_1}\right)^{\plexp}+q\sum_{i=2}^{\infty} \frac{\calHtil_i}{\calHtil_1}\left(\frac{\R_1}{\Rtil_i}\right)^{\plexp}   \Bigg| \Phibold,\boldsymbol{\mathcal{B}} \right).
\end{align}
An exact calculation of \eqref{eq:hdjkdsiu883mmsq} seems infeasible. A reasonable simplification is to obtain a lower bound approximation of $P^{(2)}_{1}$ by replacing $\frac{\calHtil_i}{\calHtil_1}$ with $\frac{\mathbb{E}[\calHtil_i]}{\mathbb{E}[\calHtil_1]}=1$. Considering this together with the fact that ${F}_{\calH_1/\calHtil_1}(x)=\frac{x}{1+x}$,  \eqref{eq:hdjkdsiu883mmsq} is approximated as follows, as also employed in \cite{10142008}:
\begin{align}
        P_1^{(2)} \approx  \frac{\R_1^{-\plexp}}{\R_1^{-\plexp} +     
        q
        \sum_{i=1}^{\infty}\left( \Rtil_i \right)^{-\plexp}}
\end{align}
Now we proceed with calculating $P_2^{(2)}$ as follows:
\begin{align}
    P_2^{(2)}&=\mathbb{P}(P_1^{(2)}> p_1\mid \Phibold)
    \notag \\
    &=\mathbb{P}\left(
    \R_1^{-\plexp}\left(1-p_1\right)>p_1 q \sum_{i=1}^{\infty}\left( \Rtil_i \right)^{-\plexp}\mid \Phibold\right)
     \notag \\
\label{eq:CS1b}
     &=
     \mathbb{P}
     \left(
     \left(\frac{\R_1}{\Rtil_1}
     \right)^{-\plexp}
     > \hat{p}_1^{\plexp} {\sum_{i=1}^{\infty}\left(\frac{ \Rtil_i}{\Rtil_1}\right)^{-\plexp}} \bigg| \Phibold
     \right)
\end{align}
Note that $\Rtil_i$ is associated with a thinned PPP with density $\zeta \lambda$. Therefore, similar to  \cite{10142008} we may leverage the approximation of substituting the term $\sum_{i=1}^{\infty}\left({ \Rtil_i}/{\Rtil_1}\right)^{-\plexp}$ in \eqref{eq:CS1b} by its expected value leveraging Lemma \ref{lm:fgsk46}. 
This reduces \eqref{eq:CS1b} to 
\begin{align}
\label{eq:hey45665}
    P_2^{(2)}
    &\approx 
    \mathbb{P} \left(\Rtil_1>\hat{p}_1\left[\frac{1+\delta\zeta}{1-\zeta}\right]^{\delta/2}\R_1 \bigg|\Phibold\right). 
\end{align}
It is observed from \eqref{eq:hey45665} that if \( \hat{p}_1 \leq \left(\frac{1 - \delta}{1 + \delta\zeta}\right)^{\delta/2} \), then \( P_2^{(2)} = 1 \). Otherwise, by comparing \eqref{eq:hey45665} with expression (b) in \eqref{eq:hjsd6835}, it follows that substituting \( \hat{p}_1 \) in the single-interferer case with \( \hat{p}_1 \left(\frac{1 + \delta\zeta}{1-\delta}\right)^{\delta/2} \) and following the same steps as in the proof of the single-interferer scenario in Theorem~\ref{th:1}, leads to \eqref{eq:hfdsnxmmzevd}.

\renewcommand{\thesection}{Appendix II} 
\section{\\ MD Reliability Calculation for Section \ref{sec:case_study_2}}
\label{sec:apxI}
In what follows, we present an analytical solution for calculating the MD reliability of the problem stated in Section \ref{sec:case_study_2}. 
Considering the presented problem statement, from \eqref{eq:MDitterative} and \eqref{eq:MDitterative_rel}, we can formulate the second-order MD reliability in a hierarchical way as follows:
\begin{subequations}
    \begin{align}
        \label{eq:THza}
        P_1^{(2)}&=\mathbb{P}
        (\mathrm{SNR}>q\mid \R,\F)
        \\
        \label{eq:THzb}
        P_2^{(2)}&=\mathbb{P}(P_1^{(2)}> p_1\mid \R)
         \\
        \label{eq:THzc}
        R_{[2]}&=\mathbb{P}(P_2^{(2)}> p_2)
    \end{align}
\end{subequations}
In the first subsection, we present the solution for the case where the available spectrum is within a monotonically increasing portion of $k(f)$. Considering that many practical THz applications exploit the
lower path loss associated with frequency bands near molecular absorption minima, in the second subsection we elaborate on the solution for the more general case where $k(f)$ is non-monotonic.
Box 2 of Fig. \ref{fig:sim3D} illustrates the frequency range corresponding to these two scenarios, wherein $f\in[\underline{f}^{(2)},\overline{f}]$ and $f\in[\underline{f}^{(1)},\overline{f}]$ correspond to the first and second scenarios respectively.

\subsection{Scenario 1: Solution Scheme if $k(f)$ is Monotonically Increasing}
\label{sec:apxIA}
 In this case, we consider that the available bandwidth corresponds to a frequency range $(\underline{f},\overline{f})$ wherein $k(f)$ is monotonically increasing. 

\subsubsection{Calculation of $P_1^{(2)}$}
\label{sec:apxIA1}
From \eqref{eq:tlTHz} and \eqref{eq:THza} we have
\begin{align}
    P_1^{(2)}
    &=\mathbb{P}\left(\mathrm{SNR}>q \mid \R,\F 
     \right)\notag
    \\
\label{eq:hdj32}
    &= 
    \mathbb{P}\left(
    \HH> c_1 q\R^2 \F^2 e^{{-k(\F) \R} }\mid \R,\F
    \right),
\end{align}
where 
\begin{align}
\label{eq:739re52}
c_1=\frac{(4\pi )^2 N_0 W}{P_{\mathrm{T}}  G_{\mathrm{T}} G_{\mathrm{R}}c^2} .     
\end{align}
The pdf of the small-scale fading is that of the Rician distribution with shape factor $K$ as follows:
\begin{align}
\label{eq:Rician}
    f_\HH(x;K) = (K+1)e^{-K-(K+1)x}I_0(\sqrt{4K(K+1)x})
\end{align}
From \eqref{eq:hdj32} and \eqref{eq:Rician}, $P_1^{(2)}$ can be obtained by calculating the ccdf of $\HH$. Following \cite{SimonBook2005}, this can be represented as follows:
\begin{align}
    P_1^{(2)}
    &=
    \int_{c_1 \R^2 \F^2 e^{{-k(\F) \R} }}^{\infty} f_\HH(x;K)dx
    \notag \\
\label{eq:364645758}
    &=    Q_1(\underbrace{\sqrt{2K}}_a,\underbrace{\F\R\sqrt{2c_1q(K+1)\exp(k(\F)\R) }}_b),
\end{align}
where $Q_1$ is the first-order Marcum Q-function. 
\subsubsection{Calculation of $P_2^{(2)}$}
\label{sec:apxIA2}
Noting that the Marcum Q-function is represented in the form of the integral of the modified Bessel function, following more analytical results in calculating $P_2^{(2)}$ and $R_{[2]}$ according to \eqref{eq:THzb} and \eqref{eq:THzc} involves the computation of multiple integrations of the modified Bessel function which is intractable using the original representation of the Marcum Q-function. To handle this, we use the exponential approximation of $Q_1(a,b)$ represented as follows \cite{6414576}:
\begin{align}
    \tilde{Q}_1(a,b) 
    &=
    \exp\left(-e^{\sum_{n=0}^M (\mu_n \ln b + \nu_n) a^n}\right)
    \notag \\
\label{eq:Marcum_approx}
    &=
    \exp\left(-e^{\nu(a) }b^{\mu(a)}\right),
\end{align}
where $\mu(a)=\sum_{n=0}^{M} \mu_n a^n$ and $\nu(a)=\sum_{n=0}^{M} \nu_n a^n$. 
Noting that $a = \sqrt{2K}$ is a fixed argument in the Marcum Q-function, we choose the coefficients $\boldsymbol{\mu} = [\mu_0, \dots, \mu_M]$ and $\boldsymbol{\nu} = [\nu_0, \dots, \nu_M]$ to minimize the least-squares (LS) error function $\mathcal{E}(a) = \int_0^\infty (Q_1(a,b) - \tilde{Q}_1(a,b))^2 \, db$. For example, over the range $a \in [1, 5]$ (i.e., $K \in [0.5, 12]$), the coefficients $\boldsymbol{\mu} = [2.174, -0.592, 0.593, -0.092, 0.005]$ and $\boldsymbol{\nu} = [-0.840, 0.327, -0.740, 0.083, -0.004]$ yield a tight approximation~\cite{6414576}. For $K = 2$, corresponding to $a = \sqrt{6}$ used in our numerical results, this gives $\mu(a) = 3.1098$ and $\nu(a) = -3.4032$. Although these values minimize $\mathcal{E}(a)$ over the full range $b \in [0, \infty)$, they may not be optimal for MD reliability calculations, where accurate approximation at specific $b$-values is more critical. In particular, since $P_2^{(2)} = \mathbb{P}(P_1^{(2)} > p_1 \mid \R) = \mathbb{E}\left[\boldsymbol{1}(Q_1(a, b) > p_1) \mid \R\right]$, it is crucial to approximate $Q_1(a, b)$ precisely at the threshold $b = b^*$, where $Q_1(a, b^*) = p_1$, as this is where the indicator function switches values. Given that the temporal target reliability $p_1$ is typically close to unity,  we have obtained the optimal values as $\mu(a)=2.4246$ and $\nu(a)=-3.3042$ for the values of $p_1$ employed in our numerical results. Leveraging the approximate representation of $Q_1$, from  \eqref{eq:THzb}, \eqref{eq:364645758} and \eqref{eq:Marcum_approx}, we can write $P_2^{(2)}$ as:
\begin{align}
    P_2^{(2)}
    &=
    \mathbb{P}(P_1^{(2)}> p_1\mid \R)
    \notag \\
    & \hspace{-10pt}
    \approx
      \mathbb{P}\left(\exp\left(-e^{\nu(a) }
     \left(
     c_2 \F\R\sqrt{\exp(k(\F)\R)}
     \right)^{\mu(a)}\right)>p_1 \mid \R\right)
     \notag \\
\label{eq:8323645005}
     &\hspace{-10pt}
     =
      \mathbb{P} \left( \F\R\sqrt{\exp(k(\F)\R)}<\tilde{p}_1 \mid \R\right),
\end{align}
where 
\begin{align}
\label{eq:p1tilde}
c_2=\sqrt{2c_1q(K+1)},\ \tilde{p}_1=\frac{1}{c_2}\times \left[-\frac{\ln(p_1)}{ e^{\nu(a)}}\right]^{1/\mu(a)}.
\end{align}
Considering \eqref{eq:8323645005}, given $\R$, let  define $\tilde{\F}(\R)$ as follows: 
\begin{align}
    \label{eq:7234045736}
    \tilde{\F}(\R)=\left\{\F\in(\underline{f},\overline{f}): \ \F \R\sqrt{\exp(k({\F})\R)}=\tilde{p}_1 \right\} 
\end{align}
Noting the monotonically increasing assumption of $k(f)$ for $f\in(\underline{f},\overline{f})$, it can be easily verified that there exists a maximum number of one solution corresponding to $\tilde{\F}(\R)$ in the desired spectrum region. We will show later that there exists exactly one solution corresponding to each desired given value of $\R$. 

Due to the non-linear representation of \eqref{eq:8323645005} as well as the non-linearity of the molecular absorption coefficient $k(\cdot)$
, it is not generally possible to write a closed-form representation of $\tilde{\F}$ in terms of $\R$. However, we will show that we may solve the problem without requiring the closed-form representation of $\tilde{\F}(\R)$.
Noting that we are studying a portion of the spectrum where $k(f)$ is a monotonically increasing function, it can be verified from \eqref{eq:7234045736} that for a given $\R$ we have
\begin{align}
    f\R\sqrt{\exp(k(f)\R)}<\tilde{p}_1, \forall f\in ( \underline{f},\tilde{\F}(\R)).
\end{align}
Considering this, together with the pdf expression of $\F$ in \eqref{eq:fF}, $P_2^{(2)}(\R)$ can be written as the cumulative distribution function (cdf) of $\F$ with input argument $\tilde{\F}(\R)$, which can be formulated $\forall m\geq 0$ 
as follows:
\begin{multline}
\label{eq:2345241990}
P_2^{(2)}(\R)=F_\F(\tilde{\F}(\R))=
    b_0+\displaystyle\sum\limits_{n=1}^{2m+1} \frac{b_n}{n} \left(\tilde{F}(\R)\right)^{n}, 
\end{multline}
where $b_n$ is the coefficient of $x^n$ in the binomial expansion of \eqref{eq:fF}, and $b_0=1-\sum_{n=1}^{2m+1}\frac{b_n}{n} (\overline{f})^{n}$ is obtained by noting $F_\F(\overline{f})=1$. For the simple case of $m=0$, corresponding to the uniform distribution of $\F$, \eqref{eq:2345241990} simplifies as follows: 
\begin{align}
    \label{eq:2345241952}
    P_2^{(2)}(\R)=
    \left[\tilde{\F}(\R)-\underline{f}\right]/(\overline{f}-\underline{f}),\ \mathrm{if}\ m=0
\end{align}
We note that the expression of $P_2^{(2)}(\R)$ in \eqref{eq:2345241990} and even in the simple case of  \eqref{eq:2345241952} is still not completely characterized, as the closed form solution of $\tilde{\F}(\R)$ is still not available.
\subsubsection{Calculation of $R_{[2]}$}
First, consider the uniform distribution of $\F$ (i.e., $m=0$). In this case,  
From \eqref{eq:THzc} and \eqref{eq:2345241952} we have $ R_{[2]}
    =
    \mathbb{P}\left({(\tilde{\F}(\R)-\underline{f})}/{(\overline{f}-\underline{f})}
    >p_2\right)
    =
    \mathbb{P}
    \left(
    \tilde{\F}(\R)>f_0
    \right)$, 
where $f_0=p_2(\overline{f}-\underline{f})+\underline{f}$. 
From \eqref{eq:7234045736} it is seen that $\tilde{\F}(\R)$ is a monotonically decreasing function of $\R$. This results in the following:
\begin{align}
\label{eq:3785gc}
    R_{[2]}=\mathbb{P}(\R<\tilde{\F}^{-1}(f_0)).
\end{align}
One can verify that \eqref{eq:3785gc} also holds for all $m\geq 0$, however for this more general case, $f_0$ can be found as the solution of the following equation:
\begin{align}
    \label{eq:hhf74836}b_0+\sum_{n=1}^{2m+1} \frac{b_n}{n} (f_0)^{n}=p_2, \ \forall m\geq 0.
\end{align}
Noting that the left side of the equality corresponds to a cdf which is a monotonically increasing function, there is a unique solution to $f_0\in[\underline{f},\overline{f}]$ which can easily be obtained using numerical methods. 
Once $f_0$ is calculated, we can compute $R_0=\tilde{\F}^{-1}(f_0)$ from \eqref{eq:7234045736} by putting $\tilde{\F}=f_0$ and finding $R_0$ as the closed form solution of
\begin{align} R_0^2{\exp(k(f_0)R_0)}=(\tilde{p}_1/f_0)^2.
\end{align}
Noting that the solution to the equation $xe^{cx}=b$ can be represented as $x=\frac{1}{c}W_0(bc)$, where $W_0$ is the principal branch of  Lambert $W$ function, after some mathematical manipulations, we obtain $R_0$ as follows:
\begin{align}
    R_0=\frac{2}{k(f_0)} W_0\left(\frac{k(f_0)\tilde{p}_1}{2f_0}\right)
\end{align}
Finally, the MD reliability is obtained as follows:
\begin{align}
\label{eq:78459346546}
     R_{[2]}
     =\int_0^{R_0} f_\R(x)dx
     &=1-\exp\left(\frac{-4\lambda \pi }{k^2(f_0)} \cdot W_0^2\left(\frac{k(f_0)\tilde{p}_1}{2f_0}\right)\right)
 \end{align}
\subsection{Scenario 2: Solution Scheme if $k(f)$ is not Monotonic}
\label{sec:apxIB}
Given that many practical THz applications exploit lower path loss associated with frequency bands near molecular absorption minima, here we consider a scenario where the channel gain $k(f)$ is non-monotonic within the frequency range $(\underline{f},\overline{f})$. Specifically, we consider the case $\underline{f}=\underline{f}^{(1)}$ illustrated in Fig. \ref{fig:sim3D}, where $k(f)$ contains a local minimum in the spanning frequency range.

\subsubsection{Calculation of $P_1^{(2)}$}
\label{sec:apxIB1}
This is achieved using \eqref{eq:364645758} as described in \ref{sec:apxIA1}.
\subsubsection{Calculation of $P_2^{(2)}$}
\label{sec:apxIB2}
Similar to the steps taken in \ref{sec:apxIA2}, $P_2^{(2)}$ is obtained from the following equation:
\begin{align}
\label{eq:83236455}
    P_2^{(2)}
    =
      \mathbb{P} \left( \underbrace{ \F\R\sqrt{\exp(k(\F)\R)}<\tilde{p}_1 }_{\mathcal{A}(\F;\R)}\mid \R\right),
\end{align}
where $\tilde{p}_1$ is given in \eqref{eq:p1tilde}. To solve \eqref{eq:83236455}, 
first we investigate the solutions of \eqref{eq:7234045736} denoted by $\F_m(\R)$ where $m$ indexes the solutions in ascending order of magnitude. Considering the behavior of $k(f)$ for $f\in[\underline{f},\overline{f}]$ where $k(\cdot)$ can initially follow a monotonically decreasing and then a monotonically increasing behavior, one can verify that we may have (a) zero, (b) one, or (c) two solution values. In what follows we investigate each case:
\begin{itemize}
    \item {\it Case (a):} If there exists no solution to  \eqref{eq:7234045736}, the event $\mathcal{A}(f;\R)$ in \eqref{eq:83236455} holds the same true/false value for all $f\in(\underline{f},\overline{f})$. Therefore, we may represent the frequency range where the corresponding event holds true as $(\tilde{\F}_1(\R),\tilde{\F}_2(\R))$, where
\begin{align}
	\tilde{\F}_1(\R)=& \underline{f} 
    \notag \\
\label{eq:8323}
		\tilde{\F}_2(\R)=& \underline{f} +(\overline{f}-\underline{f})\cdot \boldsymbol{1}(\underline{f}\R\sqrt{\exp(k(\underline{f})\R)}<\tilde{p}_1\mid \R).
\end{align}

 \item {\it Case (b):} If there exists one solution to  \eqref{eq:7234045736}, namely $\F_1$, the event $\mathcal{A}(f;\R)$ in \eqref{eq:83236455} holds same value for $f\in[\underline{f},\F_1)$ and the complemented value for $f\in[\F_1,\overline{f}]$. Hence, the frequency range where the corresponding event holds true is represented as $(\tilde{\F}_1(\R),\tilde{\F}_2(\R))$, where
   \begin{align}
	  \tilde{\F}_1(\R)=& \F_1 +(\underline{f}-\F_1)\cdot \boldsymbol{1}(\underline{f}\R\sqrt{\exp(k(\underline{f})\R)}<\tilde{p}_1 \! \mid \! \R) 
        \notag \\
\label{eq:8324}
		\tilde{\F}_2(\R)=& \overline{f} +(\F_1-\overline{f})\cdot \boldsymbol{1}(\underline{f}\R\sqrt{\exp(k(\underline{f})\R)}<\tilde{p}_1\mid\! \R).
\end{align}

\item {\it Case (c):} Finally, for the case where there exist two solutions to \eqref{eq:7234045736}, namely $\F_1$ and $\F_2$ where $\F_1\leq \F_2$, the event $\mathcal{A}(f;\R)$ in \eqref{eq:83236455} holds false for any frequency $f>\F_2$ due to the behavior of $k(.)$ corresponding to Scenario 2. Therefore, the frequency range where the corresponding event holds true is $(\tilde{\F}_1(\R),\tilde{\F}_2(\R))$, where
 \begin{align}
\label{eq:8325}
	    \tilde{\F}_1(\R)= \F_1 
       ,\
		\tilde{\F}_2(\R)= \F_2.
\end{align}
\end{itemize}
After obtaining the minimum and maximum thresholds $\tilde{\F}_1(\R)$ and $\tilde{\F}_2(\R)$, $P_2^{(2)}(\R)$ is formulated from \eqref{eq:2345241990} as
    \begin{multline}
    P_2^{(2)}(\R)=
    F_\F\left(\tilde{\F}_2(\R)\right)-F_\F\left(\tilde{\F}_1(\R)\right)=
    \\
\label{eq:43056008966}
    \displaystyle\sum\limits_{n=1}^{2m+1} 
    \frac{b_n}{n} 
    \left[ \left(\tilde{\F}_2(\R)\right)^{n}-\left(\tilde{\F}_1(\R)\right)^{n}
    \right],  \forall m \geq 0.
\end{multline}
For the simple case of the uniform distribution ($m=0$), this reduces to $P_2^{(2)}(\R)=
     [{\tilde{\F}_2(\R)-\tilde{\F}_1(\R)}]/({\overline{f}-\underline{f}})$.
\subsubsection{Calculation of $R_{[2]}$}
\label{sec:apxIB3}
\begin{algorithm}[t]
	\caption{\small\!: Calculation of the spatial-spectral-temporal MD reliability for Scenario 2}
	\begin{algorithmic}[1]
		\Statex  \hspace{-20pt} {\bf Output}: $R_{[2]}$;
		\Statex \hspace{-20pt} {\bf Initialization:}
		\State Compute $  c_1, c_2, \tilde{p}_1 $ from \eqref{eq:739re52} and  \eqref{eq:p1tilde};
        \State Let $R_{[2]}=0$, $r=0$ and $\Delta r$ be a small value;
		\Statex \hspace{-20pt} \textbf{Main Procedure:}
        \do

		\State \textbf{do}
        
		\State \hspace{7pt} Set $r=r+\Delta r$ and calculate the set of solutions of \eqref{eq:7234045736} corresponding to $\R=r$;


        


         
        \State \hspace{7pt} Calculate the set of solutions of \eqref{eq:7234045736} where $M\in \{0,1,2\}$ is the total number of solution values obtained;

    \State \hspace{7pt} Let $\tilde{\F}_1(r)$ and $\tilde{\F}_2(r)$ be obtained from \eqref{eq:8323}, \eqref{eq:8324}, or \eqref{eq:8325}, if $M=0$, $M=1$, or $M=2$ respectively.
  



        \State  \hspace{7pt} {\bf if} $\left(\sum_{n=1}^{2m+1} 
		\frac{b_{n-1}}{n} 
		\left[ \left(\tilde{\F}_2(r)\right)^{n}
        -\left(\tilde{\F}_1(r)\right)^{n}
		\right]>p_2\right)$

        \State \hspace{20pt} $R_{[2]}=R_{[2]}+ 2\pi\lambda r\exp (-\lambda \pi r^2) \Delta r$;
        
        \State  \hspace{7pt} {\bf end if}

		\State \textbf{loop until} convergence
    
	\end{algorithmic}
\end{algorithm}

Once $\tilde{\F}_1$ and $\tilde{\F}_2$ are calculated  considering any of the corresponding cases of (a), (b) and (c) elaborated in the previous part, the MD reliability can be calculated as follows:
\begin{align}
		R_{[2]}&=\mathbb{P}(P_2^{(2)}>p_2)=
        \int_0^{\infty} \boldsymbol{1}\left(P_2^{(2)}(r)>p_2\right) f_\R(r)dr
         \notag \\
		&=
        \int_0^{\infty} \boldsymbol{1}\left(\displaystyle\sum\limits_{n=1}^{2m+1} 
		\frac{b_{n-1}}{n} 
		\left[ \left(\tilde{\F}_2(r)\right)^{n}
        -\left(\tilde{\F}_1(r)\right)^{n}
		\right]>p_2 \right) \times
        \notag\\
\label{eq:5611007744465}
        &\hspace{125pt} 2\pi\lambda r\exp (-\lambda \pi r^2) dr.
\end{align}
Noting that \eqref{eq:5611007744465} can not be solved in a closed-form scheme, we present the numerical procedure for obtaining the MD reliability in Algorithm 1.  

\balance

\bibliographystyle{IEEEtran}
\bibliography{References}

@ARTICLE{8673556,
  author={Tabassum, Hina and Salehi, Mohammad and Hossain, Ekram},
  journal={IEEE Communications Surveys \& Tutorials}, 
  title={Fundamentals of Mobility-Aware Performance Characterization of Cellular Networks: A Tutorial}, 
  year={2019},
  volume={21},
  number={3},
  pages={2288-2308},
  doi={10.1109/COMST.2019.2907195}}

@techreport{3gpp_tr_38913_18,
  author = "3GPP",
  title = {{5G}; Study on scenarios and requirements for next generation access technologies (3GPP TR 38.913 version 18.0.0 Release 18)},
  type = "Technical Report (TR)",
  institution = "3GPP",
  year = "2024"
}

@ARTICLE{8474350,
  author={Deng, Na and Haenggi, Martin},
  journal={IEEE Journal on Selected Areas in Communications}, 
  title={The Energy and Rate Meta Distributions in Wirelessly Powered {D2D} Networks}, 
  year={2019},
  volume={37},
  number={2},
  pages={269-282},
  doi={10.1109/JSAC.2018.2872373}}

@ARTICLE{10142008,
  author={Gomes, André and Kibiłda, Jacek and DaSilva, Luiz A.},
  journal={IEEE Communications Letters}, 
  title={Assessing the Spectrum Needs for Network-Wide Ultra-Reliable Communication With Meta Distributions}, 
  year={2023},
  volume={27},
  number={8},
  pages={2242-2246},
  doi={10.1109/LCOMM.2023.3282140}}

@ARTICLE{9389803,
  author={Haenggi, Martin},
  journal={IEEE Communications Letters}, 
  title={Meta Distributions—{Part} 1: Definition and Examples}, 
  year={2021},
  volume={25},
  number={7},
  pages={2089-2093},
  doi={10.1109/LCOMM.2021.3069662}}

@ARTICLE{9389789,
  author={Haenggi, Martin},
  journal={IEEE Communications Letters}, 
  title={Meta Distributions—{Part} 2: Properties and Interpretations}, 
  year={2021},
  volume={25},
  number={7},
  pages={2094-2098},
  doi={10.1109/LCOMM.2021.3069681}}

@INPROCEEDINGS{8568124,
  author={Kokkoniemi, Joonas and Lehtomäki, Janne and Juntti, Markku},
  booktitle={12th European Conference on Antennas and Propagation (EuCAP 2018)}, 
  title={Simplified molecular absorption loss model for 275–400 gigahertz frequency band}, 
  year={2018},
  volume={},
  number={},
  pages={1-5},
  doi={10.1049/cp.2018.0446}}

@INPROCEEDINGS{monemi2025higherconference,
  author={Monemi, Mehdi and Rasti, Mehdi and Latva-aho, Matti and Haenggi, Martin},
  booktitle={2025 IEEE International Symposium on Personal, Indoor and Mobile Radio Communications PIMRC 2025 (accepted for publication)}, 
  title={Second-Order Meta Distribution reliability Analysis and its
Application for {UWB THz} Networks}, 
  year={},
  volume={},
  number={},
  pages={1-5},
  doi={}}

@ARTICLE{9913703,
  author={Zhou, Siyuan and Dai, Haipeng and Sun, Haihua and Tan, Guoping and Ye, Baoliu},
  journal={IEEE Transactions on Vehicular Technology}, 
  title={On the Deployment of Clustered Power Beacons in Random Wireless Powered Communication}, 
  year={2023},
  volume={72},
  number={2},
  pages={2424-2438},
  doi={10.1109/TVT.2022.3212927}}

@ARTICLE{10643015,
  author={Saeidi, Mohammad Amin and Tabassum, Hina and Alizadeh, Mehrazin},
  journal={IEEE Transactions on Wireless Communications}, 
  title={Molecular Absorption-Aware User Assignment, Spectrum, and Power Allocation in Dense {THz} Networks with Multi-Connectivity}, 
  year={2024},
  volume={23},
  number={11},
  pages={16404–16420},
  doi={10.1109/TWC.2024.3440888}}

@ARTICLE{7345601,
  author={Haenggi, Martin},
  journal={IEEE Transactions on Wireless Communications}, 
  title={The Meta Distribution of the {SIR} in {Poisson} Bipolar and Cellular Networks}, 
  year={2016},
  volume={15},
  number={4},
  pages={2577-2589},
  doi={10.1109/TWC.2015.2504983}}

@book{SimonBook2005,
  title={Digital Communication over Fading Channels},
  author={Simon, Marvin K. and Alouini, Mohamed-Slim },
  year={2005},
  publisher={ John Wiley \& Sons}
}

@ARTICLE{6414576,
  author={Bocus, Mohammud Z. and Dettmann, Carl P. and Coon, Justin P.},
  journal={IEEE Communications Letters}, 
  title={An Approximation of the First Order {Marcum Q}-Function with Application to Network Connectivity Analysis}, 
  year={2013},
  volume={17},
  number={3},
  pages={499-502},
  doi={10.1109/LCOMM.2013.011513.122462}}

@ARTICLE{10063854,
  author={Wang, Chao and Chun, Young Jin},
  journal={IEEE Access}, 
  title={Stochastic Geometric Analysis of the Terahertz ({THz})-{mmWave} Hybrid Network With Spatial Dependence}, 
  year={2023},
  volume={11},
  number={},
  pages={25063-25076},
  doi={10.1109/ACCESS.2023.3253790}}

@ARTICLE{8648502,
   author={Kalamkar, Sanket S.  and Haenggi, Martin},
  journal={IEEE Transactions on Communications}, 
  title={Simple Approximations of the {SIR} Meta Distribution in General Cellular Networks}, 
  year={2019},
  volume={67},
  number={6},
  pages={4393-4406},
  doi={10.1109/TCOMM.2019.2900676}}

@ARTICLE{8290712,
  author={Tang, Jinchuan and Chen, Gaojie and Coon, Justin P.},
  journal={IEEE Wireless Communications Letters}, 
  title={Meta Distribution of the Secrecy Rate in the Presence of Randomly Located Eavesdroppers}, 
  year={2018},
  volume={7},
  number={4},
  pages={630-633},
  doi={10.1109/LWC.2018.2805345}}

@ARTICLE{9982594,
  author={Wang, Zhe and Zheng, Jun},
  journal={IEEE Communications Letters}, 
  title={Rate Meta Distribution of Downlink Base Station Cooperation for Cellular-Connected {UAV} Networks}, 
  year={2023},
  volume={27},
  number={2},
  pages={756-760},
  doi={10.1109/LCOMM.2022.3228960}}

@ARTICLE{7893755,
  author={Salehi, Mohammad and Mohammadi, Abbas and Haenggi, Martin},
  journal={IEEE Transactions on Communications}, 
  title={Analysis of {D2D} Underlaid Cellular Networks: {SIR} Meta Distribution and Mean Local Delay}, 
  year={2017},
  volume={65},
  number={7},
  pages={2904-2916},
  doi={10.1109/TCOMM.2017.2691704}}

@ARTICLE{8115171,
  author={Cui, Qimei and Yu, Xinlei and Wang, Yuanjie and Haenggi, Martin},
  journal={IEEE Transactions on Communications}, 
  title={The {SIR} Meta Distribution in {Poisson} Cellular Networks With Base Station Cooperation}, 
  year={2018},
  volume={66},
  number={3},
  pages={1234-1249},
  doi={10.1109/TCOMM.2017.2775238}}

@ARTICLE{8587178,
  author={Salehi, Mohammad and Tabassum, Hina and Hossain, Ekram},
  journal={IEEE Transactions on Communications}, 
  title={Meta Distribution of {SIR} in Large-Scale Uplink and Downlink NOMA Networks}, 
  year={2019},
  volume={67},
  number={4},
  pages={3009-3025},
  doi={10.1109/TCOMM.2018.2889484}}

@ARTICLE{9072358,
  author={Feng, Ke and Haenggi, Martin},
  journal={IEEE Transactions on Wireless Communications}, 
  title={Separability, Asymptotics, and Applications of the {SIR} Meta Distribution in Cellular Networks}, 
  year={2020},
  volume={19},
  number={7},
  pages={4806-4816},
  doi={10.1109/TWC.2020.2987563}}

@ARTICLE{6897962,
  author={Haenggi, Martin},
  journal={IEEE Wireless Communications Letters}, 
  title={The Mean Interference-to-Signal Ratio and Its Key Role in Cellular and Amorphous Networks}, 
  year={2014},
  volume={3},
  number={6},
  pages={597-600},
  doi={10.1109/LWC.2014.2357444}}

@ARTICLE{10078093,
  author={Gomes, André and Kibiłda, Jacek and Marchetti, Nicola and DaSilva, Luiz A.},
  journal={IEEE Communications Standards Magazine}, 
  title={Dimensioning Spectrum to Support Ultra-Reliable Low-Latency Communication}, 
  year={2023},
  volume={7},
  number={1},
  pages={88-93},
  doi={10.1109/MCOMSTD.0004.2100107}}

@ARTICLE{7733098,
  author={ElSawy, Hesham and Sultan-Salem, Ahmed and Alouini, Mohamed-Slim and Win, Moe Z.},
  journal={IEEE Communications Surveys \& Tutorials}, 
  title={Modeling and Analysis of Cellular Networks Using Stochastic Geometry: A Tutorial}, 
  year={2017},
  volume={19},
  number={1},
  pages={167-203},
  doi={10.1109/COMST.2016.2624939}}

@ARTICLE{10659172,
  author={Quan, Yibo and Coupechoux, Marceau and Kélif, Jean-Marc},
  journal={IEEE Transactions on Vehicular Technology}, 
  title={Rate Meta-Distribution in Millimeter Wave {URLLC} Device-to-Device Networks With Beam Misalignment}, 
  year={2025},
  volume={74},
  number={1},
  pages={657-673},
  doi={10.1109/TVT.2024.3451487}}

@ARTICLE{8795532,
  author={Deng, Na and Haenggi, Martin},
  journal={IEEE Transactions on Communications}, 
  title={The End-to-End Performance of Rateless Codes in {Poisson} Bipolar and Cellular Networks}, 
  year={2019},
  volume={67},
  number={11},
  pages={8072-8085},
  doi={10.1109/TCOMM.2019.2934850}}

\begin{IEEEbiography}[{\includegraphics[width=1in,height=1.25in,clip,keepaspectratio]{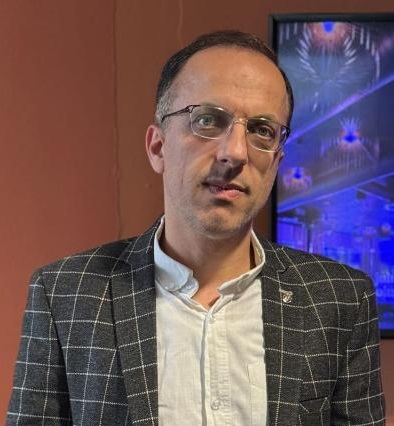}}]{Mehdi Monemi} 
    (Member, IEEE)
		received the B.Sc., M.Sc., and Ph.D. degrees all in electrical and computer engineering from Shiraz University, Shiraz, Iran, and Tarbiat Modares University, Tehran, Iran, and Shiraz University, Shiraz, Iran in 2001, 2003 and 2014 respectively. After receiving his Ph.D., he worked as a project manager in several companies and was an assistant professor in the Department of Electrical Engineering, Salman Farsi University of Kazerun, Kazerun, Iran, from 2017 to May 2023. He was a visiting researcher in the Department of Electrical and Computer Engineering, York University, Toronto, Canada from June 2019 to September 2019. He is currently an Adjunct Professor and senior researcher at the Centre
        for Wireless Communications (CWC), University of Oulu, Finland. His current research interests include resource allocation and beamforming in 5G/6G wireless networks, as well as the application of ML in various domains including machine vision and wireless networks.
\end{IEEEbiography}

\begin{IEEEbiography}[{\includegraphics[width=1in,height=1.25in,clip,keepaspectratio]{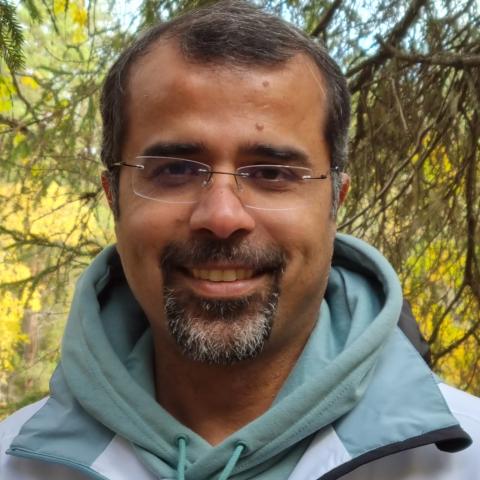}}]
    {Mehdi Rasti}
	(Senior Member, IEEE) received the B.Sc. degree in electrical engineering from Shiraz University, Shiraz, Iran, in 2001, and the M.Sc. and Ph.D. degrees from Tarbiat Modares University, Tehran, Iran, in 2003 and 2009, respectively. He is currently an Associate Professor with the Centre for Wireless Communications, University of Oulu, Finland. From 2012 to 2022, he was with the Department of Computer Engineering, Amirkabir University of Technology, Tehran. From February 2021 to January 2022, he was a Visiting Researcher with the Lappeenranta-Lahti University of Technology, Lappeenranta, Finland. From November 2007 to November 2008, he was a Visiting Researcher with the Wireless@KTH, Royal Institute of Technology, Stockholm, Sweden. 
        From June 2013 to August 2013, and from July 2014 to August 2014 he was a visiting researcher in the Department of Electrical and Computer Engineering, University of Manitoba, Winnipeg, MB, Canada. His current research interests include radio resource allocation in IoT, Beyond 5G and 6G wireless networks.	
\end{IEEEbiography}

\begin{IEEEbiography}[{\includegraphics[width=1in,height=1.25in,clip,keepaspectratio]{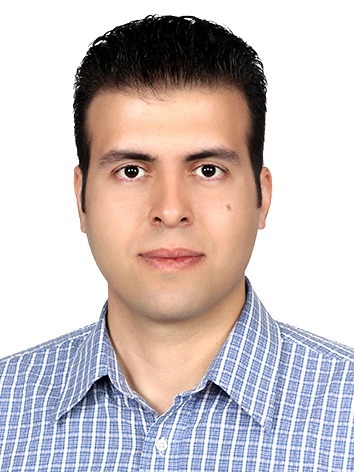}}]
    {S. Ali Mousavi} 
    received the M.Sc. degree in electrical and computer engineering from Shiraz University, Shiraz, Iran, in 2015 and Ph.D. degree in electrical and computer engineering from Shiraz University of Technology, Shiraz, Iran, in 2025. He currently serves as a lecturer in the Department of Electrical and Computer Engineering at Shiraz University, Iran. He has served as a reviewer for the IEEE Transactions on Wireless Communications. His research interest includes signal and data processing.  
\end{IEEEbiography}

\begin{IEEEbiography}[{\includegraphics[width=1in,height=1.25in,clip,keepaspectratio]{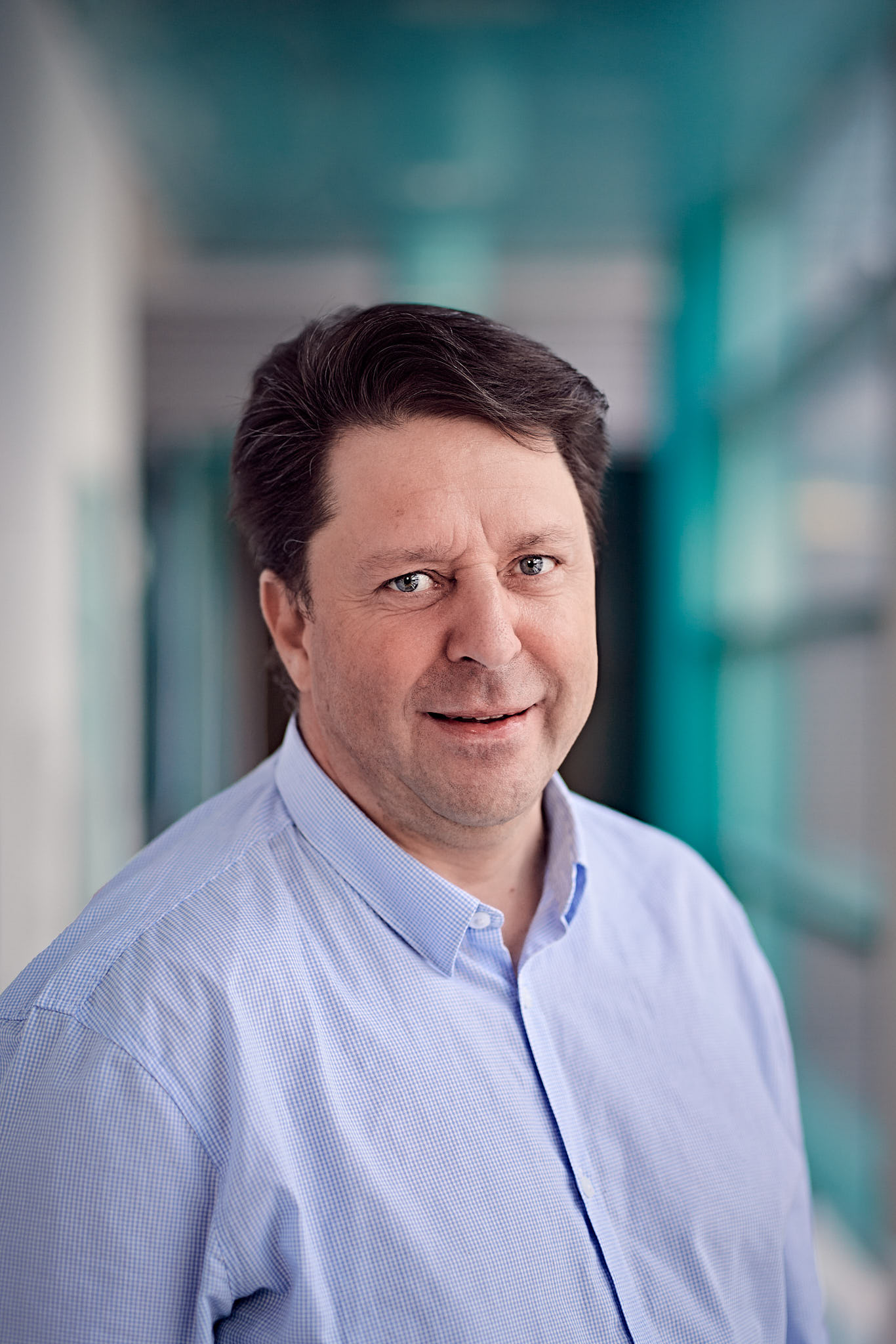}}]
    {Matti Latva-aho} (IEEE Fellow) is a distinguished expert in wireless communications. He holds M.Sc., Lic.Tech., and Dr.Tech. (Hons.) degrees in Electrical Engineering from the University of Oulu, Finland, awarded in 1992, 1996, and 1998, respectively. From 1992 to 1993, he worked as a Research Engineer at Nokia Mobile Phones in Oulu before joining the Centre for Wireless Communications (CWC) at the University of Oulu. Prof. Latva-aho served as Director of CWC from 1998 to 2006 and later as Head of the Department of Communication Engineering until August 2014. He was nominated as an Academy Professor by the Academy of Finland in 2017. He is a Professor of Wireless Communications at the University of Oulu and served as Director of the National 6G Flagship Programme. He is also a Global Fellow at The University of Tokyo. In 2025, he was appointed Vice-Rector for Research at the University of Oulu for a five-year term. With an extensive portfolio of over 600 conference and journal publications, Prof. Latva-aho has significantly advanced the field of wireless communications. His contributions were recognized in 2015 when he received the prestigious Nokia Foundation Award for his groundbreaking research in mobile communications.
        
\end{IEEEbiography}

\begin{IEEEbiography}[{\includegraphics[width=1in,height=1.25in,clip,keepaspectratio]{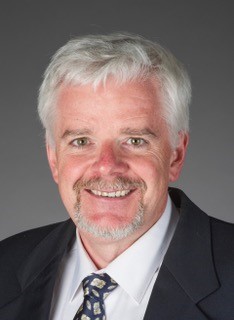}}]
    {Martin Haenggi}
    (S'95--M'99--SM'04--F'14) received the Dipl.-Ing. (M.Sc.) and Dr.sc.techn. (Ph.D.) degrees in electrical engineering from the Swiss Federal Institute of Technology in Zurich (ETHZ) in 1995 and 1999, respectively. Currently he is the Freimann Professor of Electrical Engineering and a Concurrent Professor of Applied and Computational Mathematics and Statistics at the University of Notre Dame, Indiana, USA. In 2007-2008, he was a Visiting Professor at the University of California at San Diego, in 2014-2015 he was an Invited Professor at EPFL, Switzerland, and in 2021-2022 he was a Guest Professor at ETHZ.
He is a co-author of the monographs ``Interference in Large Wireless Networks" (NOW Publishers, 2009) and ``Stochastic Geometry Analysis of Cellular Networks" (Cambridge University Press, 2018) and the author of the textbook ``Stochastic Geometry for Wireless Networks" (Cambridge, 2012) and the blog stogblog.net. His scientific interests lie in networking and wireless communications, with an emphasis on cellular, amorphous, ad hoc (including D2D and M2M), cognitive, vehicular, and wirelessly powered networks.
He served as an Associate Editor for the Elsevier Journal of Ad Hoc Networks, the IEEE Transactions on Mobile Computing (TMC), the ACM Transactions on Sensor Networks, as a Guest Editor for the IEEE Journal on Selected Areas in Communications, the IEEE Transactions on Vehicular Technology, and the EURASIP Journal on Wireless Communications and Networking, as a Steering Committee member of the TMC, and as the Chair of the Executive Editorial Committee of the IEEE Transactions on Wireless Communications (TWC). From 2017 to 2018, he was the Editor-in-Chief of the TWC.
For both his M.Sc. and Ph.D. theses, he was awarded the ETH medal. He also received a CAREER award from the U.S. National Science Foundation in 2005 and three paper awards from the IEEE Communications Society, the 2010 Best Tutorial Paper award, the 2017 Stephen O. Rice Prize paper award, and the 2017 Best Survey paper award, and he is a Clarivate Analytics Highly Cited Researcher.

\end{IEEEbiography}
    
\end{document}